%% file: holo_int_v3.tex
\documentclass[a4paper,11pt]{article}
\usepackage{jheppub}
\usepackage{amsmath,amssymb,amsfonts,graphicx,caption,subcaption,xcolor, slashed}
\usepackage[english]{babel}
\usepackage[colorlinks=true,linktocpage=true,linkcolor=blue,citecolor=blue]{hyperref}
\usepackage{marginnote}

\newcommand{\executeiffilenewer}[3]{%
\ifnum\pdfstrcmp{\pdffilemoddate{#1}}%
{\pdffilemoddate{#2}}>0%
{\immediate\write18{#3}}\fi%
}

\numberwithin{equation}{section}

\newcommand{\be}{\begin{equation}}
\newcommand{\ee}{\end{equation}}
\newcommand{\bea}{\begin{align}}
\newcommand{\eea}{\end{align}}

\newcommand{\nn}{\nonumber}

\title{Holographic charge localization at brane intersections}
\author[a]{Mario Ara\'ujo,}
\author[a]{Daniel Are\'an,}
\author[a]{Johanna Erdmenger}
\affiliation[a]{Max-Planck-Institut f\"ur Physik (Werner-Heisenberg-Institut)\\ 
F\"ohringer Ring 6, D-80805 Munich, Germany}
\author[b]{and Javier M. Lizana}
\affiliation[b]{CAFPE and Departamento de F\'isica Te\'orica y del Cosmos\\
Universidad de Granada, E-18071 Granada, Spain}
\emailAdd{maraujo@mpp.mpg.de}
\emailAdd{darean@mpp.mpg.de}
\emailAdd{jke@mpp.mpg.de}
\emailAdd{jlizan@ugr.es}

\preprint{MPP-2015-90}

\abstract{
Using gauge/gravity duality, we investigate charge localization near an interface in a strongly
coupled system. For this purpose we consider a top-down holographic
model and determine its conductivities. Our model corresponds to 
a holographic interface which localizes charge around a (1+1)-dimensional 
defect in a (2+1)-dimensional system. The setup consists of a D3/D5 
intersection at finite temperature and charge density.
We work in the probe limit, and consider massive embeddings of a D5-brane where 
the mass depends on one of the field theory spatial directions, with a profile 
interpolating between a negative and a positive value.
We compute the conductivity in the direction parallel and perpendicular to the 
interface.
For the latter case we are able to express the DC conductivity as a function of background
horizon data. At the interface, the DC conductivity in the parallel direction is 
enhanced up to  five times with respect to that in the orthogonal one. We 
study the implications of broken translation invariance for the AC and 
DC conductivities.}

\makeatletter

\renewcommand*\env@matrix[1][c]{\hskip -\arraycolsep
  \let\@ifnextchar\new@ifnextchar
  \array{*\c@MaxMatrixCols #1}}
  
\def\@eqnnum{{\normalfont\normalcolor(\theequation)}}  

\makeatother

\begin{document}
\maketitle

\section{Introduction}

The AdS/CFT correspondence \cite{Maldacena:1997re, 
Witten:1998qj,Gubser:1998bc,Aharony:1999ti} and its extensions to more
general examples of gauge/ gravity duality are by now well 
established as a powerful modeling tool for strongly coupled systems. This 
applies in particular to systems of relevance for condensed matter physics.

A number of relevant phenomena in condensed matter physics involve the 
presence of an interface between materials of different kind. Such interfaces 
represent a localized impurity which breaks translational symmetry in
the system. Broken translational symmetry allows the charge carriers
to dissipate their momentum. In the case of strong coupling where the
standard quasiparticle picture does not apply, many questions about the
exact form of this mechanism are still open.
Gauge/gravity duality reveals itself as a natural tool to further
explore momentum dissipation at strong coupling,
given that it provides a method for describing
strongly coupled systems by mapping them to weakly coupled gravity theories.

Recently, significant progress has been achieved in studying holographic 
systems with broken translation invariance by numerically solving the resulting 
equations of motion, which are in general  partial differential equations 
(PDEs). 
These include setups with different holographic realizations of lattices 
\cite{Kachru:2009xf,Kachru:2010dk,Horowitz:2012ky,Horowitz:2012gs, 
Horowitz:2013jaa, Blake:2013owa,Hartnoll:2014gaa,Donos:2014yya}
through periodically space-dependent sources,
and also 
setups implementing disordered sources 
\cite{Arean:2013mta,Arean:2014oaa,Hartnoll:2014cua,Hartnoll:2015faa}. 
Moreover, a lattice realization where PDEs are avoided, 
which goes under the name of Q-lattices given 
its resemblance to the construction
of Q-balls \cite{Coleman:1985ki}, was introduced in \cite{Donos:2013eha} and 
further explored in
\cite{Donos:2014uba,Ling:2014laa}. 
Alternatively to introducing translational 
symmetry breaking by spatially modulating the sources of 
conserved currents, momentum relaxation may also be realized by explicitly 
breaking diffeomorphism invariance in the bulk 
\cite{Andrade:2013gsa,Gouteraux:2014hca,
Mef01ford:24gia,Taylor:2014tka,Kim:2014bza,Davison:2014lua,Andrade:2014xca}, 
which in \cite{Davison:2015bea} led to progress on the study of the conductivity for systems
with broken translational symmetry.
An  example is given by helical lattices 
\cite{Iizuka:2012iv,Iizuka:2012pn,Donos:2011ff,Donos:2012gg,Donos:2012js}. 
Furthermore, translation invariance may also be broken spontaneously 
\cite{Bu:2012mq,Withers:2014sja}.

In this paper we consider the breaking of translation invariance by an interface.
We consider a top-down model involving a probe brane with a kink geometry. 
The basic idea is to incorporate the existence 
of massless modes localized on an interface
by letting the embedding vary
over one of the boundary coordinates, say $x$, in addition to being 
a function of the radial coordinate. The 
embedding function asymptotes to a positive value $m$ (with $m_q = 2 \pi \alpha' m$ the quark 
mass) for $x \rightarrow \infty$ and to $-m$ for $x \rightarrow - \infty$, while it
vanishes at $x=0$, therefore introducing a defect there.
Our work is
motivated in part by
a model constructed in \cite{HoyosBadajoz:2010ac} to holographically realize 
topological insulators by means of the D3/D7 intersection. 
This kind of configuration was used in \cite{Karch:2010mn} for both D7 and D5 probe branes to verify the expected topologically protected transport
properties for (2+1)- and (1+1)-dimensional defects, which are holographic constructions of respectively 
(2+1)-dimensional Topological Insulators and (1+1)-dimensional Quantum Spin Hall Insulators, see also \cite{Maciejko:2010tx}.
The interpretation as a topological insulator arises from the localization of 
fermions at the interface. In fact, as already shown within field theory in 
\cite{Jackiw:1975fn,Callan:1984sa}, in 3+1 dimensions there are massless 
localized fermions for a Lagrangian of the form \begin{gather} {\cal L} = \bar 
\psi \left( i \slashed{\partial} - m_q(x) \right) \psi \, , \end{gather} in which 
$m_q(x)$ jumps from a positive to a negative value at an interface. 
The thermodynamic properties of the D7-brane model with a kink \cite{HoyosBadajoz:2010ac}
were computed in \cite{Rozali:2012gf}, where the PDE equations of motion 
for the brane embedding in a black D3-brane background were solved, and the relationship between the charge density
and the chemical potential was analyzed in connection with the possible fermionic character
of the gapless interface excitations. A supersymmetric realization of the D7-brane 
holographic Topological Insulator model was given in \cite{Ammon:2012dd}.

The present paper relies on the use of probe branes, and therefore
past results for these are relevant for explaining the new structures we construct.
Hence we briefly review the pertinent features of holographic probe brane intersections.
As part of the quest for holographic models of QCD, which requires the 
presence of fundamental degrees of freedom,
probe D-brane systems were the subject of intensive study in the past. 
By considering $N_f$ D7-branes embedded in the background generated by $N$ D3-branes 
($AdS_5\times S^5$) in the limit $N_f\ll N$, one can construct the holographic dual of 
${\cal N}=4$ $SU(N)$ SYM with $N_f$ ~${\cal N}=2$ matter hypermultiplets, 
which are realized by the open strings stretching between the D3- and
D7-branes. The probe D7-branes are therefore called flavor branes
\cite{Aharony:1998xz,Karch:2002sh}. Moreover, a finite temperature is
introduced by considering the background generated by black D3-branes, while nonzero quark density , \emph{i.e.} density of the fundamental degrees of freedom, may be added by switching on the temporal component of the worldvolume gauge field on the D7-branes.
A similar construction with D5 instead of D7 flavor branes overlapping with the background D3-branes along 2+1 dimensions is dual to ${\cal N}=4$ SYM with ${\cal N}=2$ fundamental matter living on a (2+1)-dimensional defect
\cite{Karch:2000gx,DeWolfe:2001pq,Erdmenger:2002ex}. At finite density and magnetic field, these D3/D5 systems display a BKT phase transition \cite{Jensen:2010vd}.
Analyses of finite temperature setups with 
probe D7-branes \cite{Babington:2003vm,Mateos:2006nu}
and probe D5-branes \cite{Evans:2008nf} have
established two qualitatively different embeddings:
those in which the brane ends before reaching the black hole horizon, denoted Minkowski embeddings,
and those in which the brane reaches the horizon, called black hole embeddings.
There is a first order phase transition between both types of embedding which 
has been identified with the melting of mesons in the dual field theory \cite{Hoyos:2006gb}.
However, in the presence of a nonzero quark density,
only black hole embeddings are possible \cite{Kobayashi:2006sb,Mateos:2007vc,Nakamura:2007nx}.

In this work we consider a D5-brane probing the black D3-brane background. The D5-brane shares 2+1
directions with the D3-branes, and hence gives rise in the dual theory to fundamental matter living on a (2+1)-dimensional defect.
Furthermore, the embedding presents a kink-like profile as described above,
thus creating a (1+1)-dimensional interface at $x=0$, where the mass of the quarks vanishes.
A finite quark density is introduced via the temporal component of the worldvolume gauge 
field. Since our embeddings are of the black hole kind, this charge density is non-vanishing along the entire system, that is for all $x$.
There exist however homogeneous black hole embeddings where the charge density is arbitrarily small, so we can engineer kink profiles for our system such that the charge density is very small away for the interface.
We construct numerical solutions corresponding to these configurations, and check that indeed the charge density peaks at the interface.
Next, we concentrate on the study of the conductivities
of the system. We work in the linear response regime, hence to compute the conductivities we
just need study the fluctuations of the worldvolume gauge fields, which couple among themselves
and with those of the embedding field. 
Moreover, for simplicity in this work we do not consider the contribution of a WZ term proposed in \cite{Karch:2010mn} as dual to the Quantum Spin Hall effect.
This term results from  fluctuations of the RR $C_4$ form of the background.
We leave the inclusion of such a term for future work.

The study of the conductivities of our charged holographic interface gives rise to the main results of this 
work, which we now summarize.

\begin{itemize}
\item We compute the AC and DC conductivities both in the direction parallel\footnote{Since the fundamental
matter sourced by the D5 lives in a (2+1)-dimensional defect, we consider our system to be (2+1)-dimensional,
denoting by $x$ the direction orthogonal to the (1+1)-dimensional interface and by $y$ the one parallel to it. Notice that the system is
therefore homogeneous along y.}
to the interface ($\sigma^y$) and in the one orthogonal to it ($\sigma^x$). Away from the interface, both conductivities
coincide and agree with that of an homogeneous system corresponding to an embedding with constant mass
$m$. In particular, the resonances corresponding to the mesonic quasi-particles are clearly visible.
At the interface, where the quarks are massless, the conductivity exhibits at low frequency a peak reminiscent of
Drude theory. Notice that we are working in the probe approximation, and therefore the charge carriers can relax momentum into the background. Hence, no infinite DC conductivity is to be expected, see \cite{Karch:2007pd,Hartnoll:2009ns}.
  
\item Due to current conservation, $\sigma^x_{\rm DC}$ is independent of $x$ for a system 
with a codimension one impurity like ours. Following 
\cite{Iqbal:2008by,Ryu:2011vq}, we can express $\sigma^x_{\rm DC}$  purely in terms of horizon data, \emph{i.e.} the behavior of the functions describing the embedding at the black hole horizon. This DC conductivity is basically determined by the system away from the interface, where the charge density is very low.

\item
The DC conductivity is enhanced along the interface. We observe that $\sigma^y_{\rm DC}(x=0)$ is much larger than $\sigma^x_{\rm DC}$. While the latter 
is determined by the system away from the interface, where the charge density can be very low, $\sigma^y_{\rm DC}(x=0)$ is roughly proportional to the value of the charge density at the interface, and is therefore enhanced with respect to $\sigma^x_{\rm DC}$.

\item The translational symmetry breaking effects sourced by the interface 
result in an enhancement of $\sigma^y_{\rm DC}$ in its vicinity.
Although the system is homogeneous in the $y$ direction, thanks to the non-linearities of the DBI action,
a current along $y$ is sensitive to the gradients along $x$ of the embedding fields.
We observe a transfer of spectral weight in $\sigma^y$ from mid to low frequencies, resulting
in an enhancement of $\sigma^y_{\rm DC}$.

\item We study the competing effects that the presence of the interface has on the DC conductivity in the 
transverse direction, $\sigma^x_{\rm DC}$. These depend on the relative width of the interface with respect to 
the total length of the system. When that width is negligible, $\sigma^x_{\rm DC}$ is just determined by the 
homogeneous system away from the interface. On the other hand, when the interface has a sizeable width, it enhances 
$\sigma^x_{\rm DC}$. Notice that in this case the interface introduces two competing effects: an increase of 
the charge density on the one hand, and the presence of inhomogeneities along $x$ on the other.
Although, as discussed in \cite{Ryu:2011vq}, these inhomogeneities should suppress the conductivity, we  observe 
that the interface always produces an increase of $\sigma^x_{\rm DC}$ with respect to an embedding with constant 
mass $m$.
\end{itemize}

To summarize, 
up to well-known effects characteristic of holographic brane intersections, like the constant
conductivity in the high frequency limit, or the finite DC conductivity, 
the behavior of the conductivity observed in our work agrees with broad expectations for a system
where charge is localized on a (1+1)-dimensional interface. Moreover, 
some of the translational symmetry
breaking effects sourced by the interface,
as the sensitivity of $\sigma^y_{\rm DC}$ to the inhomogeneities in the orthogonal direction thanks
to the DBI action, are likely to be particular to strongly coupled systems.\footnote{See 
\cite{Rangamani:2015hka} for recent results on conductivities in the presence of
spatially modulated sources.}

This paper is organized as follows. 
Section \ref{sec:back} is devoted to the construction of the holographic interface. 
We first introduce the probe-brane embedding of interest and
write down the corresponding action. The embedding is described by two fields, for which we present the IR and UV asymptotic solutions. From the UV solutions we read the values of the chemical potential and the mass, which determine the embedding. We then characterize the part of the phase diagram accessible to the embeddings with finite charge density.
In subsection \ref{ssec:embedding} we describe the inhomogeneous embeddings that realize a charged interface. 
Finally, in subsection \ref{ssec:backres} we 
discuss the numerical methods employed to solve the equations of motion and present examples of numerical solutions corresponding to holographic interfaces. 
We study the charge density, showing that it peaks at the interface, and analyze how it scales with the chemical potential.
Section \ref{sec:conductivity} contains the main results of this work, namely the study of the conductivities
for a holographic charged interface.
We start by introducing the  fluctuations relevant to the computation of the conductivities.
We compute the quadratic action for these fluctuations, study the asymptotic solutions of the equations of motion, and discuss the relevant boundary conditions. Subsection \ref{ssec:dcsigma} is focused on the computation of the DC conductivity $\sigma^x_{\rm DC}$, which can be expressed in terms of the horizon data. Next, in subsection 
\ref{ssec:numerics}, we explain our numerical methods and boundary conditions, defining two kinds of systems, long and short,  for which $\sigma^x_{\rm DC}$ exhibits different behaviors.
Finally, in subsection \ref{ssec:results} we present and discuss the results for the conductivity of our setup.
We write our conclusions in section \ref{sec:conclusion}, where we also discuss some possible directions of future research. 
%
%
We have furthermore included two appendices. Appendix \ref{app:eoms} contains the background equations of motion.
In appendix \ref{app:action}
we write down the quadratic action for the fluctuations relevant for studying the conductivities.

\section{Localized charge at brane intersections}
\label{sec:back}
In this section we consider D3/D5 intersections at nonzero temperature and finite charge density, 
namely in the presence of a finite density of the fundamental matter dual to the open strings stretching 
between the D3- and D5-branes.
The supersymmetric intersection of $N$ D3- and $N_f$ D5-branes along 2+1 spacetime dimensions
is dual to (3+1)-dimensional ${\cal N}=4$ SYM with $N_f$ fundamental hypermultiplets living on a 
(2+1)-dimensional defect \cite{Karch:2000gx,DeWolfe:2001pq}.
We work in the probe limit and at nonzero temperature, hence we treat the D5-branes as probes in the geometry 
generated by the black D3-branes.

\subsection{Black D3-branes}
According to the AdS/CFT prescription originally proposed in  
\cite{Maldacena:1997re},  ~${\cal N}=4$ super Yang-Mills theory with an $SU(N_c)$ 
gauge group is holographically dual to type IIB string theory on $AdS_5\times 
S^5$ with $N_c$ units of flux of the Ramond-Ramond five form. The string coupling 
$g_s$,
and the coupling of the gauge theory $g_{\rm YM}$ 
are related through
$g_s=g^2_{YM}/2\pi$. 
The AdS curvature radius $L$ is furthermore related to $N_c$ and the string 
tension $(2\pi \alpha')^{-1}$ by $L^4/ \alpha^2=2 g^2_{YM} N_c\equiv 2\lambda$, where $\lambda$ is the
't Hooft coupling. 
In the limit of large $N_c$ and large $\lambda$ the string side of the duality
reduces to weakly coupled classical gravity.

We are interested in finite temperature configurations, and therefore consider
the geometry generated by $N_c$ black D3-branes, whose metric reads\footnote{
This metric is related to the more standard Schwarzschild-AdS presentation via the
change of coordinates $z^2=2L^4/\left(u^2+\sqrt{u^4-u_0^4}\right)$.}
\cite{Mateos:2006nu}
\begin{equation}
 ds^2=\frac{L^2}{z^2}\left(-\frac{f(z)^2}{h(z)}\,dt^2+h(z)\,d\vec{x}^2+dz^2
\right)+L^2\,d\Omega_5^2\,,
\label{eq:metric}
\end{equation}
where $\vec{x}=(x_1,x_2,x_3)$, $d\Omega_5^2$ is the metric of a unit radius $S^5$, and
\begin{equation}
 f(z)=1-\frac{z^4}{z_0^4}\,,\qquad h(z)=1+\frac{z^4}{z_0^4}\,.
\end{equation}
This geometry becomes asymptotic to $AdS_5\times S^5$ in
the small $z$ limit, with the boundary of $AdS_5$ being at $z=0$, while it presents a horizon
at $z=z_0$. Accordingly, the Hawking temperature of the black hole reads
\begin{equation}
 T={\sqrt{2}\over \pi\,z_0}\,.
 \label{eq:temp}
\end{equation}
It is useful to define the dimensionless coordinates
\begin{equation}
 (\tilde z, \tilde x_\mu) = {1\over z_0} (z,x_\mu)\,,
 \label{eq:coordredef}
\end{equation}
in terms of which the metric takes the form \eqref{eq:metric} with $z_0 = 1$. From now on we always
use these dimensionless coordinates and drop the tilde for presentational purposes.

\subsection{Probe D5-brane}
As anticipated, we embed a probe D5-brane in the background generated by the 
stack of D3-branes \eqref{eq:metric}.
The embedding is better described by writing the metric of the $S^5$ in terms of two $S^2$ as follows
\begin{equation} 
d\Omega_5^2=d\theta^2+\sin^2\theta\,d\Omega_2^2+\cos^2\theta\,d\tilde\Omega_2^2\,.
\label{s5metric}
\end{equation}
We consider the following configuration
\begin{equation}
 \begin{tabular}{r c c c c c c c c}
 {} & $t$ & $x_1$ & $x_2$ & $x_3$ & $z$ & $\Omega_2$ & $\tilde\Omega_2$ & 
$\theta$ \\ 
  D3 &	$\times$	&	$\times$	&	$\times$	&	
$\times$	&	{}	&	{}	&	{}	& {} \\
  D5 &	$\times$	&	$\times$	&	$\times$	&	
{}	&	$\times$	&	$\times$	&	{}	&	
{}\\
 \end{tabular}
 \label{eq:embtable}
 \end{equation}
where the D5-brane shares two Minkowski directions with the D3-branes generating
the background, is extended along the radial direction $z$, and wraps an $S^2$ ($\Omega_2$)
inside the $S^5$, while it is located at a fixed point of the remaining $S^2$ ($\tilde\Omega_2$).
The embedding is then described by the coordinate $\theta$, which determines
the radius of the $S^2$ wrapped by the D5-brane. In order to simplify the analysis
we define
\begin{equation}
\cos \theta = \chi\,,
\end{equation}
in terms of which we describe the embedding.

We are interested in configurations with finite charge density 
of the fundamental fields introduced by the flavor D5-brane. Therefore
we turn on a nonzero temporal component of the $U(1)$ worldvolume gauge field
\begin{equation}
 A= A_t\, dt\,.
\end{equation}
Indeed, its boundary value determines the value of the chemical potential in the dual theory.
The embedding of the probe D5-brane is then determined by the two fields $\chi$ and $A_t$.

\subsubsection{Action and equations of motion}
\label{section:action}
In order to describe setups where the embedding of the D5-brane depends on one of the spatial
directions, we let the fields $\chi$ and $A_t$ depend on the radial variable $z$, 
and on one of the Minkowski directions; $x_1$ (just $x$ in the following).\footnote{As 
far as the dependence on $x_2$ and $x_3$ is concerned, 
it is consistent to locate the D5-brane
at $x_3=0$, and consider embeddings homogeneous along $x_2$, which we call $y$ in the following.}
Hence the embedding is characterized by 
$\chi(z,x)$ and $A_t(z,x)$.

The dynamics of the system is governed, in the probe approximation, by the DBI 
action for the D5-brane in the background sourced by the black D3-branes
\begin{equation}
S=-N_f T_{D5} \int d^6x\,\sqrt{-\det(P[g]+2\pi\,\alpha'\,F)}\,,
\label{dbi}
\end{equation}
where $P[g]$ is the pullback of the metric on the worldvolume of the D5-brane and
$F$ the field strength of the worldvolume $U(1)$ gauge field.
For our setup the DBI action can be written as\footnote{It is straightforward to check
that for the embedding at hand there is no contribution to the action coming from the WZ term.}
\be
S=-N_f\,T_{D5}\,L^6
\int dt\,d^2x\,dz\,d\Omega_2\,f\,z^{-4}\,
\sqrt{h\, (1-\chi^2)\,(S_\chi+S_\phi + S_{\text{int}})}\,,
\label{eq:action}
\ee
with
\bea
 & S_{\chi}=1-\chi^2+z^2 \chi'^2+\frac{z^2\,\dot\chi^2}{h}\,,\\
 & S_\phi=-\frac{z^4(1-\chi^2)}{f^2}\left(h\,\phi'^2+\dot \phi^2\right)\,,\\
 & S_{\text{int}}=-\frac{z^6(\dot{\chi}\phi'-\chi' \dot{\phi})^2}{f^2 }\,,
\end{align}
where a tilde denotes a derivative with respect to $z$ and a dot  a 
derivative with respect to $x$.
Moreover, $\phi$ stands for the dimensionless temporal component of the gauge field defined via
\begin{equation}
\phi=2\pi\alpha'\, {z_0\over L^2}\, {A_t}\,.
\label{eq:phidef}
\end{equation}

The equations of motion for $\phi(z,x)$ and $\chi(z,x)$ can be readily obtained
from the action \eqref{eq:action}. The resulting lengthy expressions are shown in appendix \ref{app:eoms}.
In the following we analyze their IR ($z\to1$) and UV ($z\rightarrow 0$) asymptotic
solutions.
\subsubsection*{IR Asymptotics}
As we explain below, we are interested in solutions describing black hole embeddings for
which the brane ends at the horizon. Hence regularity at the horizon requires
$\phi$ and $\chi'$ to vanish there, restricting the IR solution to the following form
\begin{subequations}
\begin{align}
  & \phi(z,x)=a^{(2)}(x)\,(1-z)^2+{\cal O}((1-z)^{3})\,, \\
 &\chi(z,x)=C^{(0)}(x)+C^{(2)}(x)\,(1-z)^2+{\cal O}((1-z)^{3})\,,
 \end{align}
 \label{eq:bhexpansion}
\end{subequations}
with 
\begin{equation}
 C^{(2)}(x)=\frac{\left(2-a^{(2)}(x)^2\right) \left[C^{(0)''}(x)(C^{(0)}(x)^2-1)
-C^{(0)}(x) \left(3
   C^{(0)'}(x)^2+4\right)+4 C^{(0)}(x)^3\right]}{8 \left[C^{(0)'}(x)^2-2 C^{(0)}(x)^2+2\right]}\,.
\end{equation}

\subsubsection*{UV Asymptotics}
At the boundary, the asymptotic form of the fields $\chi(z,x)$ and $A(z,x)$ can 
be found to be of the form 
\begin{subequations}
\begin{align} 
& \phi(z,x)=\mu(x)-\rho(x) z+{\cal O}(z^{2})\,, \label{eq:phiUV}\\
&\chi(z,x)=m(x) \,z+\psi(x)z^2+{\cal O}(z^{3})\,. \label{eq:chiUV}
\end{align}
\label{eq:backUV}
\end{subequations}
Let us recall that the holographic dictionary relates $\mu(x)$, and $\rho(x)$ respectively
to the chemical potential and charge density of the $U(1)$ flavor symmetry supported
by the D5-brane. As for the asymptotic form of $\chi$, the leading piece $m(x)$
is proportional to the  asymptotic distance $\bar M$ between the probe D5 and the D3-branes generating
the background, and is therefore interpreted as the quark mass. The subleading contribution
$\psi(x)$ is associated to the vacuum expectation value of the bilinear quark-antiquark operator
sourced by the flavor D5-brane, namely the quark condensate.
In order to make these identifications more precise, let us bring the dimensions back into the game.
First, we recall that the dimensionful chemical potential $\bar\mu$ and charge density
$ \bar \rho$ are read from the asymptotic form of the temporal component
of the gauge field as
 \be
 A_t = \bar\mu(x) - \bar\rho(x) z+\dots\,.
 \ee
Hence, recalling the  redefinitions \eqref{eq:coordredef} and \eqref{eq:phidef},
the temperature of the black hole \eqref{eq:temp},
 and using that $\sqrt{\lambda}=L^2/(\sqrt{2}\alpha')$
 we arrive at
 \be
 \mu = {2\over\sqrt{\lambda}}\,{\bar\mu\over T}\,,\qquad
 \rho = {2\sqrt{2}\over\pi\sqrt{\lambda}}\, {\bar \rho \over T^2}\,.
 \label{eq:murhodef}
 \ee
 As for the quark mass, we follow \cite{Mateos:2006nu} and define $M_q = \sqrt{\lambda}\,\bar M/2$,
 where $\bar M = m/z_0$. This allows us to write
 \be
 m={2\sqrt{2}\over\pi\sqrt{\lambda}}\,{M_q\over T}\,.
 \label{eq:mdef}
 \ee

As is clear from \eqref{eq:backUV}, the UV solutions depend on four  parameters (functions of $x$):
$\mu$, $\rho$, $m$, and $\psi$,  while as we see in \eqref{eq:bhexpansion},
the IR behavior is determined by two free functions, namely $C^{(0)}(x)$ and $a^{(2)}(x)$. Thus
we expect a two-parameter family of solutions, which we choose to describe in terms
of the chemical potential $\mu$, which we choose to be independent of $x$,
and the mass $m(x)$.
Consequently, once $\mu$ and $m(x)$ are fixed, the embedding of the probe D5-brane is completely
fixed.

\subsection{Inhomogeneous embeddings and charge localization}
\label{ssec:embedding}
The phenomenology of Dp/Dq brane intersections, both at zero and finite temperature, with 
and without charge
density, has been thoroughly studied over the last ten years, see for
instance \cite{Arean:2006pk,Myers:2006qr,Erdmenger:2007cm} 
and references therein.

We now review some features that are relevant for our construction.
Embeddings of the probe D5-brane in the black D3-brane background generally fall into 
two qualitatively different categories labeled Minkowski and black hole (BH).
Minkowski embeddings are those in which 
the D5-brane never reaches the black hole, while for the BH embeddings the probe brane ends at the horizon.
Minkowski embeddings exist above a certain value of the mass, while BH embeddings are possible for any mass.
A phase transition between these two embeddings occurs as the ratio $M_q/ T$ is varied; this is known
as meson melting \cite{Mateos:2006nu} since stable mesonic states exist for Minkowski embeddings, but not
for black hole ones.
This phase transition also happens in the presence of a chemical potential, and the corresponding phase diagram
has been studied in \cite{Evans:2008nf} (see also \cite{Mateos:2007vc}).

\begin{figure}
 \centering
 \def\svgwidth{0.8\columnwidth}
\executeiffilenewer{phasediag.svg}{phasediag.pdf}%
{inkscape -z -D --file=phasediag.svg %
--export-pdf=phasediag.pdf --export-latex}%
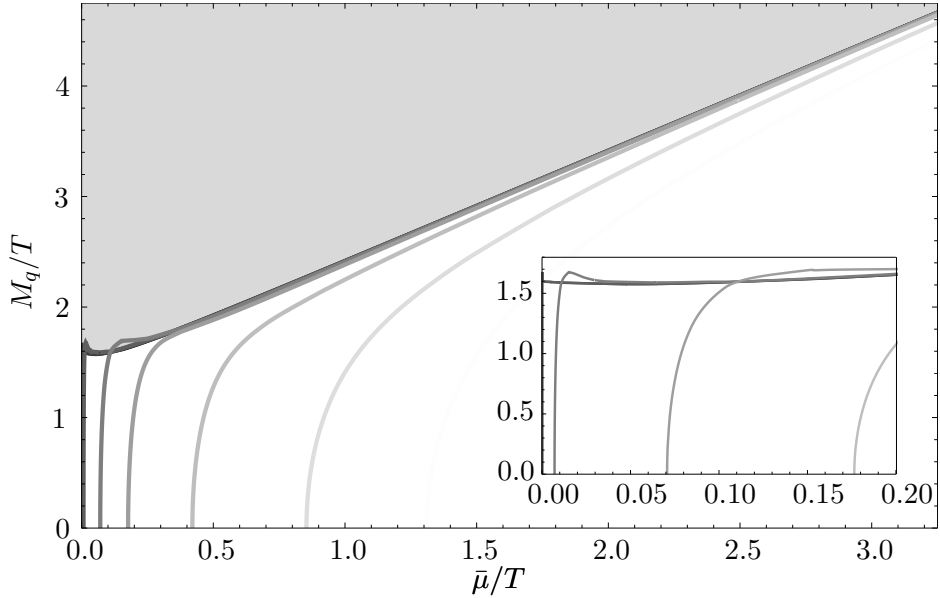%

\caption{Quark mass $m=M_q/T$ versus chemical potential $\mu=\bar\mu/T$ for black hole embeddings corresponding
to various values of the charge density $\rho=\bar\rho/T^2$, decreasing from right to left: $\rho= 2$,
1.25, 0.6, 0.25, 0.1, 0.01, $10^{-4}$, $10^{-6}$. The shaded area is not accessible by BH embeddings,
and is asymptotically delimited (at large $\mu$) by $m>1.00\mu+1.41$. In the inlay we zoom in on the region
of low $\mu$.}
\label{fig:BHphase}
\end{figure}

\begin{figure}
 \centering
 \def\svgwidth{0.75\columnwidth}
\executeiffilenewer{phasediag2.svg}{phasediag2.pdf}%
{inkscape -z -D --file=phasediag2.svg %
--export-pdf=phasediag2.pdf --export-latex}%
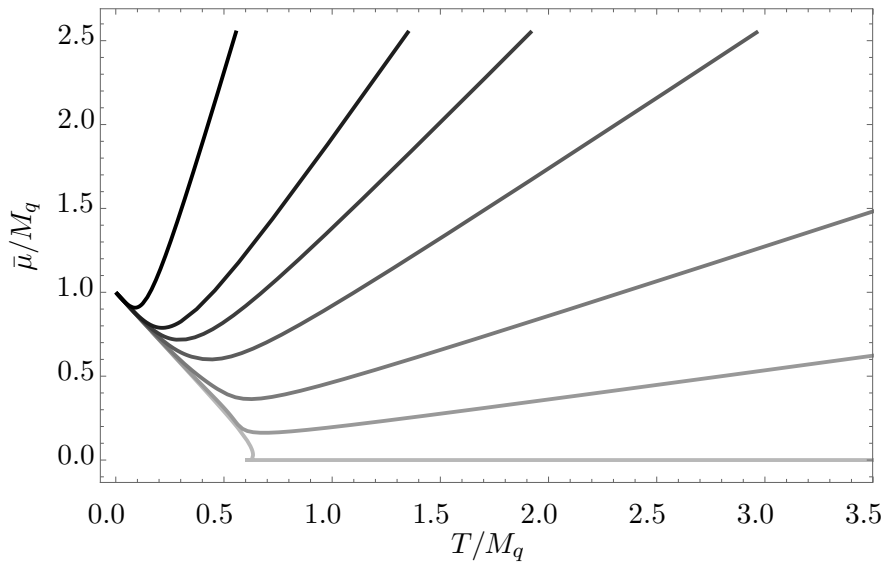%

        \caption{As in fig. \ref{fig:BHphase} we plot lines of constant charge density for BH embeddings.
        From bottom (light gray) to top (black) they correspond to
        $\rho= 10^{-6},  0.25, 0.6, 1.25, 2, 3, 10$.}
\label{fig:phasediag2}
\end{figure}

A feature that is crucial for our construction stems from the fact that at nonzero charge density
only BH embeddings are possible. As explained in \cite{Kobayashi:2006sb},
the fundamental strings realizing the charge density would have nowhere to end in a Minkowski embedding.
Moreover, as illustrated in figure \ref{fig:BHphase}, for a large enough $M_q/ T$, BH embeddings exist only above
a nonzero chemical potential $\bar\mu/T$. In particular, the shaded region in the plot is only accessible
by Minkowski embeddings, which have zero charge density. 
Since it will be useful later, in figure \ref{fig:phasediag2} we plot the same data as in fig. \ref{fig:BHphase}, 
but now  for $\bar\mu/M_q$ versus $T/M_q$. In terms of these variables the forbidden region corresponds to the 
triangular region visible at  $T/M_q\lesssim0.6$

As explained above, our purpose is to construct an embedding depending on one spatial direction $x$, that
localizes charge density along an interface situated at $x=0$. This can be done, following
\cite{HoyosBadajoz:2010ac}, by means of an embedding with constant chemical potential $\mu=\bar\mu/T$, 
and an $x$-dependent mass ($m=M_q/T$)
\begin{equation}
 m(x)=M\left(\frac{2}{1+e^{-a\,x}}-1\right)\,,
 \label{eq:massprofile}
\end{equation}
that interpolates between two constant values; $M$ at $x\to-\infty$, and $-M$ at $x\to \infty$, 
while vanishing at the origin, $m(0)=0$.\footnote{As explained in
\cite{Karch:2010mn}, embeddings with positive and negative $m$ correspond respectively to the D5-brane sitting at
opposite poles of the $\tilde\Omega_2$ in \eqref{eq:embtable}.
As in \cite{HoyosBadajoz:2010ac} we account for the case $m<0$ by letting $\chi$ become negative.}
Notice that $a$ is a constant parameter that fixes the steepness of the kink. 
Ideally, if we chose $\mu$ and $M$ to lie in the shaded region of figure \ref{fig:BHphase},
asymptotically, at $x\to\pm\infty$, the embedding should be of the Minkowski type,
and therefore the charge density would vanish. At the interface ($x=0$) though, the mass vanishes,
the brane must intersect the BH, and, for a nonzero $\mu$, some charge density is induced.
In such a construction the charge density would exactly vanish towards the spatial edges, while it would peak 
at the interface. Notice that by increasing $a$ 
in \eqref{eq:massprofile}  the transition can be made as abrupt as desired, and therefore
the charge may be localized at $x\sim0$.
However, such an embedding, with a varying topology along $x$, turns out to be too challenging in numerical terms.
Instead, we do as in \cite{Rozali:2012gf}, and settle for a more modest construction: we choose $M$
and $\mu$ to be just outside, but at the edge of, the shaded region of figure \ref{fig:BHphase}. 
So we deal with embeddings that are of the BH kind everywhere.
Notice that in principle one can pick $M$ and $\mu$ such that the corresponding embedding has an arbitrarily
small charge density induced at the edges. Then, effectively the charge density is localized
around the interface, where the embedding becomes massless.

\subsection{Numerical Results}
\label{ssec:backres}
In this section we present numerical solutions
realizing the inhomogeneous embeddings described in the previous section.
In order to construct those embeddings
we must solve the equations of motion for the fields $\chi(z,x)$ and $\phi(z,x)$,
and the inhomogeneous mass profile boundary condition \eqref{eq:massprofile} implies
that we have to deal with two coupled second order PDEs which we
solve numerically.

Before describing our numerical method, let us recall the boundary conditions to be
imposed on the two equations of motion. First, in the UV ($z=0$), from \eqref{eq:backUV}
and \eqref{eq:massprofile}, we impose
\be
\chi'(0,x)=M\left(\frac{2}{1+e^{-ax}}-1\right)\,,\qquad
\phi(0,x)= \mu\,,
\label{eq:backUVimp}
\ee
where $\mu$ determines the homogeneous chemical potential of the solution, while
$M$ fixes the mass of the embedding at the edges of the system.
On the other hand, at the horizon ($z=1$) the asymptotic solutions \eqref{eq:bhexpansion}
result in the following boundary conditions
\be
\phi(1,x)=0\,,\qquad
\chi'(1,x)=0\,.
\label{eq:backIRimp}
\ee

As for the boundary conditions at the spatial edges, note that the symmetry of our setup does not allow 
the use of periodic boundary conditions. We take our system to have a finite length ($x\in[-L,L]$), but require it to be large
enough so that it resembles an homogeneous embedding towards the spatial edges. Consequently
we impose the following Neumann boundary conditions
\be
\dot\chi(z,\pm L)=0\,,\qquad
\dot\phi(z,\pm L)=0\,,
\label{eq:backspBC}
\ee
which ensure that the effects of the inhomogeneity sourced by the mass profile \eqref{eq:massprofile}
fade away towards the edges.

Regarding the numerics, we resort to pseudospectral methods implemented in Mathematica,
discretizing the plane $(z,x)$ on a grid of Chebyshev points, and then solving 
the resulting set of nonlinear algebraic equations via
Newton-Raphson iteration. Defining the variations of the fields $f=(\chi, \phi)$ in each iteration
by $\delta f$, we consider the accuracy of our solution to be given by $\text{Max }|\delta f|$.

In addition, we can benefit from the symmetry of our setup
by noting that $\chi(z,x)$ is an odd function of $x$, whereas $\phi(z,x)$ is even
\begin{equation}
\chi(z,x)=-\chi(z,-x)\,,\qquad
\phi(z,x)=\phi(z,-x).
\label{eq:parity}
\end{equation}
This follows from the form of the equations of motion
together with our UV boundary conditions \eqref{eq:backUVimp},
and helps us making the numerics more efficient in two ways.
First, it allows us to solve for half the range along $x$, imposing
\eqref{eq:backspBC} at $x=L$, while in view of \ref{eq:parity} at $x=0$ we must have
\be
\chi(z,0)=0\,,\qquad
\dot \phi(z,0)=0.
\label{eq:backBCx0}
\ee
Second, given that Chebyshev collocation points are more
densely concentrated towards the boundaries of the interval,
this reduction of the integration range results in a better
accuracy of our solutions around the interface (at $x=0$),
where the gradients along $x$ are larger.

Having described the numerical method we employ to construct the embeddings,
we now present the output of the numerical computation.
We have used a grid of $50\times 50$ collocation points for the  half-interval of 
integration ranging from $x=0$ to $x=10$.
%
%
\begin{figure}
\centering
\def\svgwidth{0.9\columnwidth}
\executeiffilenewer{afieldb.svg}{afieldb.pdf}%
{inkscape -z -D --file=afieldb.svg %
--export-pdf=afieldb.pdf --export-latex}%
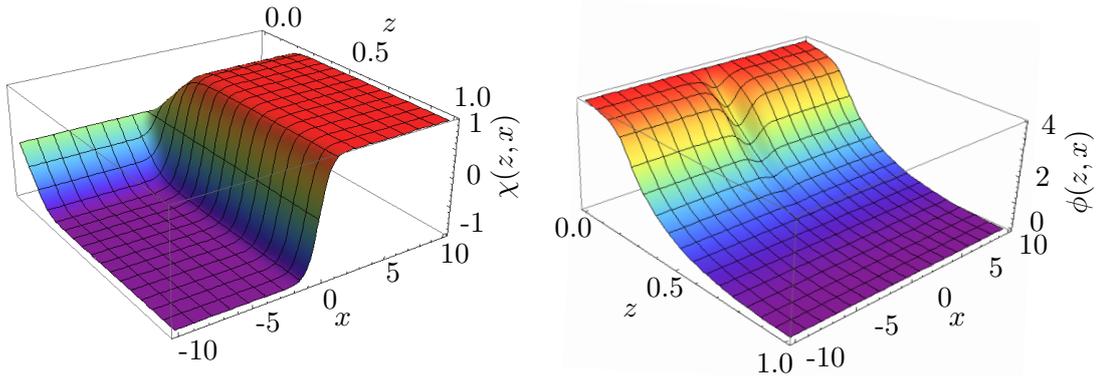%

\caption{Computed solutions of $\chi(z,x)$ and $A(z,x)$ for 
$\mu=4$ and $M=5.34$.
}
\label{fig:background}
\end{figure}

According to what we have discussed in section \ref{ssec:embedding} we have chosen the values
of $M$ and $\mu$ to be such that the charge density induced at the edges of our
system is much lower than that at the interface. 
In figure \ref{fig:background} we plot the numeric solution 
for the fields $\chi(z,x)$ and $\phi(z,x)$, for a case where
$\mu=4$ and $M=5.34$. The parameter $a$ is chosen so that the numerics remain stable while
still having a steep embedding. $a=3$ turns out to be good enough for this\footnote{While the numerics 
allow for
much larger values of $a$ for the computation of the background, these pose some difficulties when 
it comes
to solving for the perturbation fields.}.
It is worth mentioning how the spatial inhomogeneity introduced
by the step-like boundary condition \eqref{eq:massprofile} affects differently the two
fields defining our setup. While for $\chi$ the inhomogeneity is amplified
towards the horizon, for the gauge field $\phi$ it dies away towards the horizon.

\begin{figure}
\centering
\def\svgwidth{0.6\columnwidth}
\executeiffilenewer{chargedensity.svg}{chargedensity.pdf}%
{inkscape -z -D --file=chargedensity.svg %
--export-pdf=chargedensity.pdf --export-latex}%
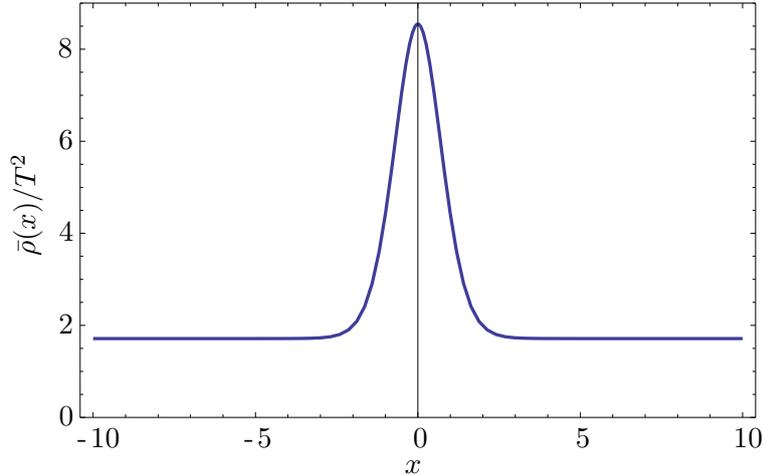%

\caption{Charge density $\rho(x)=\bar\rho(x)/T^2$ for the background in figure 
\ref{fig:background}. The base line is at $\rho=1.71$, and the peak reaches $\rho=8.54$.}
\label{fig:chargedensity}
\end{figure}

In figure \ref{fig:chargedensity}
we present the resulting charge density, which through \eqref{eq:phiUV} is given
by the radial derivative of $\phi$ evaluated at the boundary. Indeed, we see that the
charge density peaks at the interface, where its value is about five times the value at the edges.

\begin{figure}
 \centering
 \def\svgwidth{0.6\columnwidth}
\executeiffilenewer{moshe.svg}{moshe.pdf}%
{inkscape -z -D --file=moshe.svg %
--export-pdf=moshe.pdf --export-latex}%
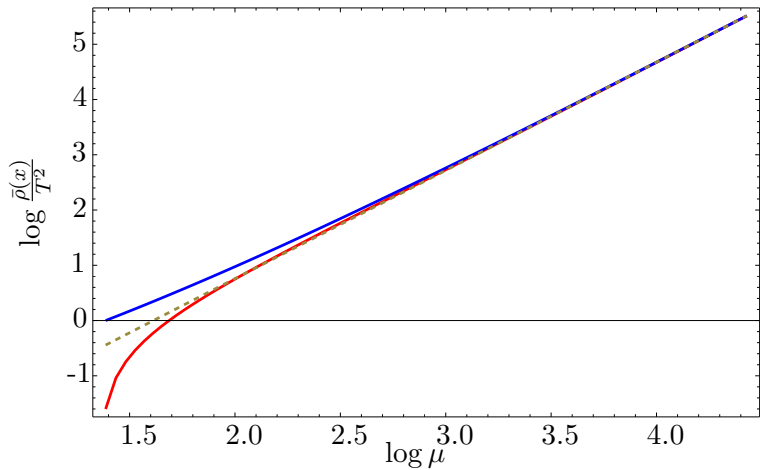%

        \caption{Charge density $\rho(x)$ versus chemical 
potential $\mu$ for an embedding with $M=5.3$. The blue line corresponds to the interface,
while the red one to one of the spatial edges. 
The charge density has been normalized to unity at $\mu=1, \,m=0$. 
The dashed line illustrates the fit 
$\rho = \mu^{1.96}$ performed for $\mu>74$.}
\label{fig:moshewrong}
\end{figure}

Finally, it is interesting to study how the charge density depends on the chemical potential, both at the
interface and far from it. This is plotted in figure \ref{fig:moshewrong}, where we observe that 
the scaling $\rho\propto\mu^2$ expected for a D3/D5 intersection \cite{Nogueira:2011sx} is approached 
everywhere in our system for large enough $\mu$.

\section{Conductivities}
\label{sec:conductivity}
After constructing a holographic setup localizing charge along a (1+1)-dimensional defect, we go on to study its response to an applied electric field. In the rest of this work we
study the AC and DC conductivities of our system both in the direction parallel, $y$, and orthogonal to the defect,
$x$.
In order to do so, we compute the linear response of our background when an electric
field is switched on along the boundary. We must then
study the fluctuations of the worldvolume gauge field along the Minkowski directions.
In general these are coupled among them and to the fluctuations of the embedding field $\chi$, thus we have
to solve for the whole set of coupled fields.

\subsection{Fluctuations}
\label{ssec:flucts}
We are interested in switching on fluctuations of the gauge field realizing an electric field 
of constant modulus and frequency $\omega$ along the boundary, hence at $z=0$ those fluctuations must satisfy
\be
f_{ti} = \left(i\omega\, \mathfrak{e}_i\right)e^{i\omega\,t}\,,\qquad (i=x,y)\,.
\label{eq:efield}
\ee
where $f_{ti}$ stands for the field strength of the fluctuations of the gauge field, and we are
considering both the case when the electric field is orthogonal to the interface (with 
constant modulus $\mathfrak{e}_x$), and parallel to it ($\mathfrak{e}_y$).
Consequently, we must study the following set of fluctuations\footnote{We are working with the 
dimensionless coordinates \eqref{eq:coordredef}, hence $\omega$ is dimensionless, and in terms of 
the dimensionful frequency $\mathfrak{w}$, one has $\omega=\sqrt{2}\, \mathfrak{w}/(\pi\,T)$.}
\begin{subequations}
\begin{align}
& A_\mu = A_\mu (z,x) + a_\mu (z,x)e^{i\omega t}\,,\label{eq:amufluc}\\
& \chi =\chi (z,x) + c(z,x)e^{i\omega t}\,,\label{eq:chifluc}
\end{align}
\label{eq:flucdef}
\end{subequations}
where the uppercase $A_\mu$ and $\chi$ stand for the background fields, while lowercase $a_\mu$ and $c$ refer
to the fluctuations of the gauge field and the embedding scalar respectively\footnote{We need apply the 
same field redefinition as in \eqref{eq:phidef} to the fluctuations of the gauge field. However, for 
notational simplicity we do not introduce a new label and just assume that we have rescaled the fields as 
$a_\mu\to L^2/(2\pi\alpha'\,z_0)\,a_\mu\,$.
}. Our background,
described by $A_t(z,x)$ and $\chi(z,x)$, is both time invariant and translation invariant along the $y$ direction.
This allows us to Fourier transform both along $t$ and $y$,
and since we are interested only in the conductivity we consider our fluctuations to have no net spatial
momentum. 
In addition, we choose to work in the radial gauge and therefore set
\be
a_z(z,x)=0\,.
\label{eq:radialgauge}
\ee

We are working in the linear response regime, hence the equations of motion for the fluctuations 
\eqref{eq:flucdef} follow from expanding the DBI action \eqref{dbi} up to second order in those fluctuations. 
The resulting quadratic action $S^{(2)}$ is shown in appendix \ref{app:action}.
Although straightforward to derive, the resulting equations of motion are lengthy and we do
not reproduce them here.
However, it is worth mentioning that the component of the gauge field fluctuations parallel to the defect, $a_y$,
decouples from the rest of the fluctuations. Hence to study the conductivity $\sigma_y$ we only
need to solve the corresponding linear partial differential equation (PDE) for $a_y$.
Instead, $a_x$ is coupled to both $a_t$, and $c$. Their dynamics is described by a system of three second 
order linear PDEs plus a first order constraint PDE resulting from the equation of motion for $a_z$.

The conductivity is given by  the retarded correlator of the worldvolume $U(1)$ current. In order to compute it, we must solve the equations of motion of the fluctuations with infalling boundary conditions at the horizon. 
Then we can read the electric current on the boundary, $j_i$, and therefore compute the conductivity as\footnote{Here and in the following we are rescaling the conductivity by the  dimensionless constant appearing in front of the action \eqref{eq:action}, \emph{i.e.} $\sigma\to\sigma/(N_f T_{D5}\,L^6)$, so that we recover the usual high frequency CFT result $\lim_{\omega\to\infty}\sigma=1$.}
\begin{align}
 \sigma_i(\omega,x)=\frac{j_i}{i\omega\,\mathfrak{e_i}}=\lim_{z \to 0} 
\frac{f_{iz}}{f_{ti}}\,,\qquad (i=x,y)\,.
\label{eq:sigmadef}
\end{align}

We now study the asymptotic behavior of the fluctuations both in the IR and UV. This allows us
to properly choose the boundary conditions when solving the corresponding equations of motion.
It is easy to check that the fields behave in the UV ($z\rightarrow 0$) as
\begin{subequations}
\begin{align}
 &a_\mu(z,x) = a_\mu^{({\rm b})}(x) - j_\mu(x)\,z + {\cal O}(z^2)\,,\qquad (\mu=t,x,y)\,,\label{eq:amuUV}
 \\
 &c(z,x)=c^{({\rm b})}(x)\,z+ {\cal O}(z^2)\,,
\end{align}
\label{eq:flucUV}
\end{subequations}
while in the IR  ($z\rightarrow 1$) they take the form
\begin{subequations}
\begin{align}
& a_\mu = (1-z)^{i \alpha_\pm} \left( a_\mu^{(0)}(x)+ a_\mu^{(1)}(x)\,(1-z)+{\cal O}((1-z)^2)\right)\,, 
\label{eq:amuUVbc}\\
& c = (1-z)^{i \alpha_\pm} \left( c^{(0)}(x)+ c^{(1)}(x)\,(1-z)+{\cal O}((1-z)^2)\right)\,,
\label{eq:cUV}
\end{align}
\label{eq:flucIR}
\end{subequations}
with 
\be
\alpha_\pm=\pm {\omega\over 2\sqrt{2}}\,.
\ee
Notice that with our conventions (see \eqref{eq:flucdef}) it is the positive root $\alpha_+$
the one corresponding to an infalling solution at the horizon. For this choice we obtain the following solution
for the first order ($x$-dependent) coefficients
\be
c^{(1)}=\frac{i\omega}{4 \sqrt{2}}\,c^{(0)}\,,\qquad
a_t^{(0)}=0\,,\qquad 
a_i^{(1)}=\frac{i\omega}{4 \sqrt{2}}\,a_i^{(0)}\,; \qquad (i=x,y)\,, \label{eq:IRbc1}\\
\ee
while higher order coefficients are determined in terms of these.

We now have all ingredients needed to compute the conductivity by solving the equations of motion
of the fluctuations.
First, as is clear from the form of the IR solutions \eqref{eq:flucIR}, it is useful to redefine the fields as
\be
\tilde a_\mu(z,x) = (1-z)^{-i\alpha_+}\,a_\mu(z,x)\,,\qquad
\tilde c(z,x) = (1-z)^{-i\alpha_+}\,c(z,x)\,.
\label{eq:fieldredef}
\ee
Then, at the horizon we impose the following mixed Dirichlet and Robin boundary conditions
\be
\tilde a_t(1,x) = 0\,,\qquad \tilde a_i'(1,x)=\frac{i\omega}{4 \sqrt{2}}\,\tilde a_i(1,x)
\,,\qquad \tilde c'(1,x)=\frac{i\omega}{4 \sqrt{2}}\,\tilde c(1,x)\,.
\label{eq:irbcs}
\ee
At the boundary we want our fluctuations to source an homogeneous electric field \eqref{eq:efield}.
When computing the conductivity parallel to the defect (we need only solve for $a_y$), we impose
\be
a_y (0,x) = 1\,,
\label{eq:ayUVBC}
\ee
where we are normalizing the modulus of the electric field to unity ($\mathfrak{e}_y=1$).
Instead, to compute $\sigma_x$ we must solve for $a_x$, $a_t$, and $c$. Again, we want our fluctuations
to source solely an electric field in the $x$ direction. Hence
\begin{subequations}
\begin{align}
&c'(0,x)=0\,,\label{eq:cUVBC}\\
&a_x(0,x)-{1\over i\omega}\partial_x a_t(0,x)=1\,,
\label{eq:axatUVBC}
\end{align}
\label{eq:amuandcUVBC}
\end{subequations}
where the first condition ensures that no fluctuations of the mass are sourced, whereas the second
implies that an homogeneous electric field along $x$, normalized to $\mathfrak{e}_x=1$,
is turned on at the boundary. In addition, in the UV we impose the fulfillment of the constraint equation,
which reduces to
\be
i \omega\, \partial_z a_t(0,x)- \partial_x \partial_z a_x(0,x)=0\,.
\label{eq:constUV}
\ee
Notice that in terms of the asymptotic solutions \eqref{eq:amuUV} this boundary condition is nothing
else than the conservation of current
\be
\partial_t\left(e^{i\omega\,t} j_t(x)\right)-e^{i\omega\,t}\,\partial_x j_x(x)=0\,.
\label{eq:currentcons}
\ee
As expected, it is straightforward to check that the partial derivative along $z$ of the constraint
equation vanishes for solutions of the equations of motion. This ensures that the constraint is satisfied
for all $z$ by any solution of the equations of motion that obeys the constraint
on a constant $z$ slice.

Finally, one should notice that although we have allowed for both $a_t$ and $a_x$ to be nonzero at the boundary,
one can proceed as in \cite{Donos:2013eha, Donos:2014yya} and apply a gauge transformation 
$e^{i\omega t}\,\Lambda(z,x)$ that brings the boundary field configuration to
%
\begin{equation}
  a_\mu(0,x)\,e^{i\omega\,t}\,dx^\mu \rightarrow \left(a_\mu + \partial_\mu \Lambda(0,x)\right) e^{i\omega t}\,
  dx^\mu= e^{i\omega\,t} dx\,,
\label{eq:amugauget}
\end{equation}
which makes clear that the only boundary source is that corresponding to an homogeneous electric field,
and other nonzero sources are just gauge artifacts.\footnote{Notice that one can 
always choose $\Lambda(z,x)$ such that it vanishes at the horizon (so $a_t$ is still
zero there), and also satisfies $\partial_z \Lambda(0,x)=0$ (and hence $a_z(0,x)=0$).}.

Summing up, in order to compute the conductivity $\sigma_y(\omega)$ we must solve the equation of motion
of $a_y$ with boundary conditions \eqref{eq:irbcs} and \eqref{eq:ayUVBC}, and then read
the conductivity from \eqref{eq:sigmadef}. On the other hand, to calculate $\sigma_x$ we need solve
the equations of motion of $a_x$, $a_t$, and $c$, imposing \eqref{eq:irbcs} at the horizon, and
(\ref{eq:amuandcUVBC} - \ref{eq:constUV}) on the boundary; and again read $\sigma_x(\omega)$ from \eqref{eq:sigmadef}.
We will discuss the boundary conditions at the spatial boundaries when describing our numerical methods.
But before that, in the next section we analyze the DC limit of the conductivity, and show
how $\sigma_x^{\rm DC}$ can be computed from the background horizon data, with no need to solve for the fluctuations.

\subsection{DC conductivity}
\label{ssec:dcsigma}
We can follow the procedure of \cite{Iqbal:2008by} (as applied for instance in \cite{Ryu:2011vq} to a DBI action)
and compute the DC conductivity along the direction perpendicular to the interface $\sigma_x^{\rm DC}$
in terms of the background functions evaluated at the horizon.

Let us start by defining the radial coordinate $\zeta$ through
\begin{equation}
d\zeta =\sqrt{\frac{h(z)}{f(z)^2}}\,dz\,,
\label{eq:efcoord}
\end{equation}
and note that the horizon (at $z=1$) is located at $\zeta=\infty$ in the new coordinate.
When expressed in this coordinate, the equations of motion for the fluctuation 
fields $a_x(z,x)$ and $a_z(z,x)$ in the DC limit, $\omega \rightarrow 0$, respectively take the form
\begin{equation}
\partial_\zeta \left( {\cal F}(\zeta,x)\, \partial_\zeta a_x \right)= 0\,,
\qquad
\partial_x \left( {\cal F}(\zeta,x)\, \partial_\zeta a_x \right)= 0\,,
\label{eq:dceqs}
\end{equation}
where we have defined
\begin{equation}
{\cal F}(z,x)= f \left(1-\chi^2\right)^{3/2} \sqrt{\frac{h}{\Gamma}}\,,
\label{eq:Fdef}
\end{equation}
with
\begin{equation}
\begin{split} 
\Gamma &= -z^4 h \left\{\phi'^2 \left[h(1-\chi^2)+z^2\, \dot\chi^2\right]
-2 z^2 \phi'\,\dot\phi\, \chi' \, \dot\chi+\dot\phi^2 (1-\chi^2+z^2\, \chi'^2)\right\} \\
 &\quad - f^2 \left[ h\left(\chi^2-1-z^2\, \chi'^2\right)-z^2\,\dot\chi^2\right].\\
\end{split}
\end{equation}
The equations of motion \eqref{eq:dceqs} ensure that the combination ${\cal F}\, \partial_\zeta a_x$
is a constant. It is thanks to the existence of this conserved quantity that one can
express $\sigma_x^{\rm DC}$ in terms of the background functions evaluated at the horizon.
First, notice that at the boundary
\begin{equation}
 {\cal F}(0,x)=1,
 \label{eq:ff1}
\end{equation}
and let us define the function
\begin{equation}
X(\zeta,x)=-\frac{\partial_\zeta a_x(\zeta,x)}{a_x(\zeta,x)},
\label{eq:defX}
\end{equation}
which at the horizon satisfies
\begin{equation}
X(\infty,x)=i\omega\,,
\label{eq:inwaveX}
\end{equation}
due to the ingoing wave boundary condition imposed on $a_x$.
In terms of $X(\zeta,x)$, the conductivity from \eqref{eq:sigmadef} reads
\be
\sigma_x(\omega,x)={X(0,x)\over i\omega}\,a_x(0,x)\,,
\label{eq:sigmaxX}
\ee
where we have normalized the modulus of the electric field to one.
Next, in the DC limit we can expand $X$ in a power series in $\omega$ as
\be
X(\zeta,x)= i \omega \,a(\zeta,x)+{\cal O}(\omega^2)\,,
\label{eq:Xexp}
\ee
and $a_x$ and $a_t$ at the boundary as
\begin{subequations}
\begin{align}
& a_t(0,x)= i \omega \,p(x)+{\cal O}(\omega^2)\,,\\
& a_x(0,x)= 1+\partial_x p(x)+{\cal O}(\omega)\,,
\end{align}
\label{eq:ataxexp}
\end{subequations}
where $a(\zeta,x)$ and $p(x)$ are fixed by the equations of motion \eqref{eq:dceqs}. Moreover,
notice that \eqref{eq:ataxexp} is such that the condition \eqref{eq:efield} of having a constant
electric field (with $\mathfrak{e}_x=1$) at the boundary is automatically satisfied. 
At the horizon, the ingoing wave condition \eqref{eq:inwaveX} translates into
\be
a(\infty,x)=1\,.
\label{eq:aIRc}
\ee
Plugging the expansions (\ref{eq:Xexp}, \ref{eq:ataxexp}) into eq. \eqref{eq:sigmaxX} we obtain
\be
\sigma_x^{\rm DC}=a(0,x)(1+\partial_x p(x))\,.
\label{eq:sigdcx}
\ee
Using the definition \eqref{eq:defX} together with the expansions 
(\ref{eq:Xexp}, \ref{eq:ataxexp}), the equations of motion \eqref{eq:dceqs} imply
that
\begin{equation}
 {\cal F}(\zeta,x) \,a(\zeta,x)\, (1+\partial_x p(x)) 
\label{eq:dccons}
\end{equation}
is a constant\footnote{We have taken into account that $a_x(\zeta,x)=a_x(0,x)+{\cal O}(\omega)$.}.
Now, notice that \eqref{eq:dccons} when evaluated at $\zeta=0$  reduces precisely to the expression
\eqref{eq:sigdcx} for the conductivity.
Hence we conclude that $\sigma_x^{\rm DC}$ is a constant. By evaluating \eqref{eq:dccons} at the horizon
we arrive at the following expression for the DC conductivity
\be
\sigma_x^{\rm DC}={\cal F}(z=1,x)\, (1+\partial_x p(x))\,,
\label{eq:sigdcdef1}
\ee
which is indeed a constant as required by current conservation. Notice though, that this expression for $\sigma_x^{\rm DC}$ still depends on the fluctuations
through the field $p(x)$ which should in principle be determined by solving the corresponding equations of motion.
However, this dependence can be eliminated and $\sigma_x^{\rm DC}$ expressed solely in terms
of the background horizon data. Integrating the expression \eqref{eq:sigdcdef1} over the whole sample we can write
\be
\sigma_x^{\rm DC}\,{1\over 2L}\int_{-L}^{L}\frac{dx}{{\cal F}(1,x)}={1\over 2L}\int_{-L}^{L}dx\,(1+\partial_x p(x))\,,
\ee
and assuming the condition
\be
{1\over2L}\,\int_{-L}^{L}dx\,(1+\partial_x p(x))=1\,,
\label{eq:pbc}
\ee
(to be justified below) we arrive at the following expression for the DC conductivity
\be
\sigma_x^{\rm DC}={2L\over \int_{-L}^{L}\frac{dx}{{\cal F}(1,x)}}\,,
\label{eq:sigdcdef2}
\ee
which allows us to calculate $\sigma_x^{\rm DC}$ purely in terms of background functions evaluated
at the horizon. In terms of the IR asymptotic solutions for $\phi$ and $\chi$,
given in eq. \eqref{eq:bhexpansion}, ${\cal F}(z=1,x)$ can be written as 
\begin{equation}
 {\cal F}(z=1,x)=\frac{2 
\left({C^{(0)}(x)}^2-1\right)^{3/2}}{\sqrt{\left(a^{(2)}(x)^2-2\right) 
\left(2-2 {C^{(0)}(x)}^2+{C^{(0)'}(x)}^2\right)}}.
\label{eq:sigmadexpansion}
\end{equation}

To end this section, let us discuss the condition \eqref{eq:pbc} that must be satisfied by the fluctuations.
First, notice that when rewritten in terms of $a_t(z,x)$ it boils down to
\be
\int_{-L}^{L}dx\, \partial_x a_t(0,x) = 0\,.
\label{eq:atintbc}
\ee
This would be automatically satisfied for a periodic system, but also in a setup like ours if we assume
that the system is long enough for the effects of the interface to fade away towards the edges of the sample.
As we discuss below, it can be checked that in that case the solution
for the fluctuations 
asymptotes towards the edges to that of a homogeneous system, for which  $a_t=0$ and then 
\eqref{eq:atintbc}  holds.
A more general argument for requiring \eqref{eq:atintbc} to hold is as follows.
Notice that even though we allow $a_t$ to be nonzero at the boundary, as illustrated by \eqref{eq:amugauget}
our configuration is gauge equivalent to one where $a_x$ is the only source at the boundary 
\cite{Donos:2013eha, Donos:2014yya}.
Then, $a_t(0,x)$ is pure gauge,  \emph{i.e.} $\Lambda(0,x)$, and gauge invariance of the action in presence of a
conserved current implies that$\int_{-L}^{L}dx\, \partial_x \Lambda(0,x) = 0$, which justifies the assumption
\eqref{eq:atintbc}.


\subsection{Numerics}
\label{ssec:numerics}
As discussed above, in order to compute the conductivity, we solve the equations of motion of the fluctuations
numerically. We now briefly describe the numerical methods employed, and specify the boundary conditions
imposed at the spatial edges of the system.\\
We solve the equations of motion of the fluctuation fields \eqref{eq:fieldredef} on the same Chebyshev grid used 
for the background. To simplify the numerics, we make use of
the parity along $x$ of the fields in our problem, namely
\begin{equation}
\begin{split}
 & a_t(z,x)=-a_t(z,-x)\,,\qquad  a_x(z,x)=a_x(z,-x)\,,\\
 & c(z,x)=c(z,-x)\,,\qquad  a_y(z,x)=a_y(z,-x)\,,
\end{split}
\end{equation}
which follows straightforwardly from the (linear) equations of motion taking into account that
the background fields satisfy \eqref{eq:parity}.
As in the case of the background, this allows us to actually solve for half the system, between $x=-L$ and $x=0$, and given the distribution of
points in a Chebyshev grid, greatly increases the resolution close to the interface, where the gradients in $x$ are larger.
The IR and UV boundary conditions are given by eqs. (\ref{eq:IRbc1}, and \ref{eq:ayUVBC} - \ref{eq:constUV})
as discussed above.

Regarding the boundary conditions at the spatial edges ($x=\pm L$), again, periodic boundary conditions cannot be used
due to the symmetry of the problem. Let us focus first on the the three coupled fields $a_t$, $a_x$, and $c$, which allow us to compute $\sigma_x$.
By studying the asymptotic form of the coupled PDEs at the spatial boundaries, one can show that a solution
is completely determined once the values of $a_t$ and $c$, or those of their derivatives $\partial_x a_t$
and $\partial_x c$, are fixed at each spatial boundary\footnote{Solving 
the system asymptotically at one edge, once the the values of $a_t$ and $c$, and their derivatives
$\partial_x a_t$ and $\partial_x c$, are fixed, the asymptotic solution for $a_x$, $a_t$ and $c$ is fully
determined (one needs plug in the UV and IR boundary conditions too). 
Then, to fully determine a solution of the system one can fix the values of $a_t$ and $c$, and their derivatives,
at one edge, or equivalently impose two conditions at one edge and two more at the other. We consider fixing 
$a_t$ and $c$ both at $x=L$ and $x=-L$, or, alternatively, fixing $\partial_x a_t$ and $\partial_x c$ at $x=\pm L$.}.
The case of $a_y$ is simpler, for we just need to solve a linear PDE, and as spatial boundary conditions we either
fix the value of the function, or of its derivative $\partial_x a_y$, at the spatial boundaries.
Both when computing $\sigma_x$ and $\sigma_y$ we consider two different sets of boundary conditions 
as we now describe.
\subsubsection *{Damping boundary conditions. Long systems}
A reasonable boundary condition is derived from the assumption that the system is long enough for all the inhomogeneities sourced by
the interface to die away towards the edges, that is from the requirement that the fluctuations become independent of $x$ there, namely
\be
\partial_x a_t(z,\pm L)=0\,, \qquad
\partial_x c(z,\pm L)=0\,,
\label{eq:atcdamp}
\ee
while $a_x$ is left free as discussed above.

To compute the conductivity in the direction parallel to the interface ($\sigma_y$) we only need to solve for $a_y$. The damping boundary condition at the spatial boundaries is then
\be
\partial_x a_y(z,\pm L)=0\,.
\label{eq:aydamp}
\ee

\subsubsection *{Boundary conditions. Short systems}
One can instead be interested in a situation in which the system is not long enough for the inhomogeneities of the fluctuations to vanish 
towards
the spatial boundaries. 
Consequently, a relaxation of the boundary conditions considered above for long systems would consist in allowing
the fluctuations to have a nonzero derivative at the spatial boundary. Then, we must impose
Dirichlet boundary conditions there.
As discussed in the previous section, $a_t$, which is odd, must obey \eqref{eq:atintbc}, hence
an alternative boundary condition that allows  for a nonzero $\partial_x a_t$ at the boundary is
\be
a_t(z,\pm L)=0\,.
\label{eq:atbcx}
\ee
Analogously we require $c$ to also vanish at the boundaries
\be
c(z,\pm L)=0\,.
\label{eq:cbcx}
\ee
Notice that these boundary conditions are nothing else than the requirement that $a_t$ and $c$ reach the solution of the homogeneous problem exactly at the edge. We should bear in mind that when computing the conductivity 
$\sigma_x$ of an homogeneous system, only $a_x$ has to be turned on, hence $c$ and $a_t$, which decouple from $a_x$, 
vanish identically. It is straightforward to check, both analytically and numerically, that these boundary conditions are satisfied whenever the previous more restricting damping boundary conditions are imposed. Yet the opposite is not true, and for short enough systems the solutions are such that $\partial_x a_t$ and $\partial_x c$ are non-vanishing at the spatial boundaries.

Finally, for $a_y$ one can also consider a Dirichlet boundary condition which requires that $a_y$ reach
the homogeneous solution at the boundary, namely the solution for $a_y$ in a homogeneous background 
characterized by the values of the chemical potential and the mass far away from the interface
\be
a_y(z,\pm L)=a_y^{\rm hom}(z)\,.
\label{eq:cbcx}
\ee

Except for when we specifically focus on long systems (fig. \ref{fig:comparesigmadcS}), 
in the rest of this work we consider our systems to be short, and consequently impose the
boundary conditions above.
In particular, we set $L=10$ (remember $x\in [-L,L]$, with the interface located at $x=0$),
and  fix $a=3$ in \eqref{eq:massprofile}, as discussed in section \ref{ssec:backres}.
Moreover, as for the background, we use grids of size $50\times 50$, except for the results plotted
in figures \ref{fig:sigmaACyvshomog} and \ref{fig:rhoeffect}
which were obtained with a grid of size $N_z\times N_x = 50\times 35$.

\subsection{Results}
\label{ssec:results}
Finally, in this section we present our results  for the conductivities of a holographic
interface.
After a glance at the optical conductivity for the entire range of the coordinate $x$, we focus on 
its behavior at the interface. We end the section by studying the DC conductivity.
\begin{figure}
\centering
\def\svgwidth{0.9\columnwidth}
\executeiffilenewer{sigmawxXandY.svg}{sigmawxXandY.pdf}%
{inkscape -z -D --file=sigmawxXandY.svg %
--export-pdf=sigmawxXandY.pdf --export-latex}%
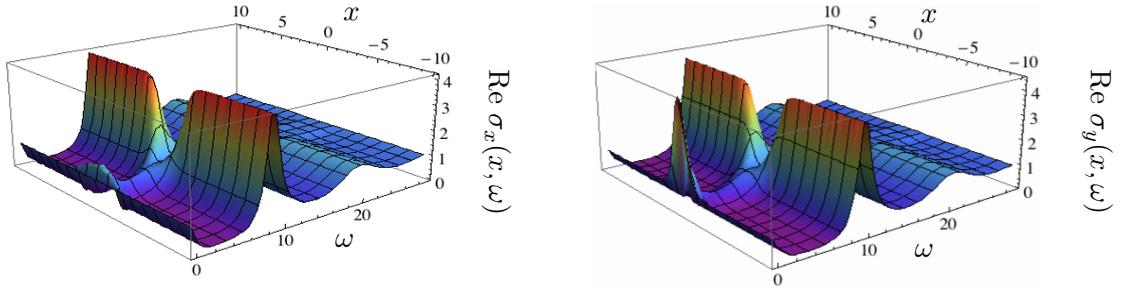%

\caption{Real optical conductivities as a function of frequency and position, for a background with $\mu=4$ and 
$m=5.3$. Notice that the main differences between $\sigma_x$ and $\sigma_y$ occur at low frequencies, and
close to the interface ($x=0$).}
\label{fig:sigma3d}
\end{figure}

In figure \ref{fig:sigma3d} we plot the real part of the optical conductivities $\sigma_x$ and $\sigma_y$ as
functions of the frequency $\omega$, and the position $x$ for a background with $\mu=4$
and\footnote{We remind the reader that the value of $M$ sets the mass reached by the inhomogeneous embedding 
\eqref{eq:massprofile} at the edges, away from the interface.} 
$M=5.3$.
At intermediate and large frequencies, and away from the interface, both conductivities
are very similar, not only to each other, but also to the conductivity of the equivalent homogeneous system, 
\emph{i.e.} the one given by an homogeneous embedding with the same values of mass and chemical potential that
characterize our system away from the interface. In particular, we observe the presence of the resonances
given by the quasi-normal modes corresponding to the `melting' vector
mesons. In fact, as found in \cite{Erdmenger:2007ja},
the effective meson masses, which correspond to peaks in the spectral
function, are in one-to-one relation with the  
frequencies 
\begin{equation}
 \omega_{\rm res}= m \sqrt{2(k+1)(k+2)}\,, \qquad k=0,1,2,\dots\, ,
 \label{eq:resonances}
\end{equation}
which correspond to the masses of stable mesons \cite{Kruczenski:2003be}. 
Away from the interface the system becomes homogeneous, and therefore
we expect the conductivity to reproduce the homogeneous result. 
On the other hand, near the interface the conductivities in the directions parallel and orthogonal to it 
behave quite differently, presenting interesting features which we elucidate below.

\begin{figure}
\centering
\def\svgwidth{\columnwidth}
\executeiffilenewer{sigmaACyvshomogALL.svg}{sigmaACyvshomogALL.pdf}%
{inkscape -z -D --file=sigmaACyvshomogALL.svg %
--export-pdf=sigmaACyvshomogALL.pdf --export-latex}%
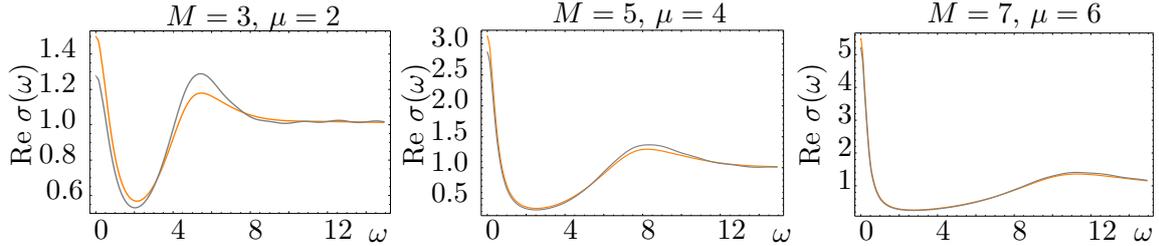%

\caption{Plots of the real part of $\sigma_y(\omega)$ (orange lines) at a point $x=x_0=0.749$ in the 
vicinity of the interface for different pairs of values $(M, \mu)$. The gray line stands for the equivalent 
homogeneous conductivity $\sigma_y^{\rm h}(\omega)$, obtained for an homogeneous system with the same mass 
$m=m(x_0)$, and chemical potential as our setup at that point (from left to right $m(x_0)=2.4262$, $4.044$, 
and $5.661$ respectively).
}
\label{fig:sigmaACyvshomog}
\end{figure}

Let us first study the effects of the inhomogeneities on the conductivity along the direction parallel 
to the interface.
For a system like ours,  homogeneous along the $y$ direction, one could naively expect
that at each point $x$, the conductivity in the $y$ direction, $\sigma_y(\omega,x)$, be very similar 
to that of an homogeneous system having the same mass and chemical potential as our setup at that point,
which we denote $\sigma_y^{\rm h}$.
However, as we discuss below, 
$\sigma_y(\omega,x)$ is sensitive to the spatial gradients of the inhomogeneous embedding, and therefore
becomes different from $\sigma_y^{\rm h}$ where the spatial gradients are large.
To illustrate this, in figure \ref{fig:sigmaACyvshomog}
we plot $\sigma_y(\omega)$ and  $\sigma_y^{\rm h}(\omega)$ at the point where 
the difference between them is maximal, which is of course close to the interface.
Interestingly, with respect to the equivalent homogeneous case we observe a transfer of spectral weight 
from intermediate to very low frequencies resulting in a larger DC conductivity in the presence of the 
interface. Moreover, we have checked that the relative enhancement increases with decreasing $\mu$ for a 
given $M$ as one moves toward the phase transition in the phase diagram \ref{fig:BHphase}.

\begin{figure}
\centering
\def\svgwidth{\columnwidth}
\executeiffilenewer{sigmaACyvshomogGRADIENTS.svg}{sigmaACyvshomogGRADIENTS.pdf}%
{inkscape -z -D --file=sigmaACyvshomogGRADIENTS.svg %
--export-pdf=sigmaACyvshomogGRADIENTS.pdf --export-latex}%
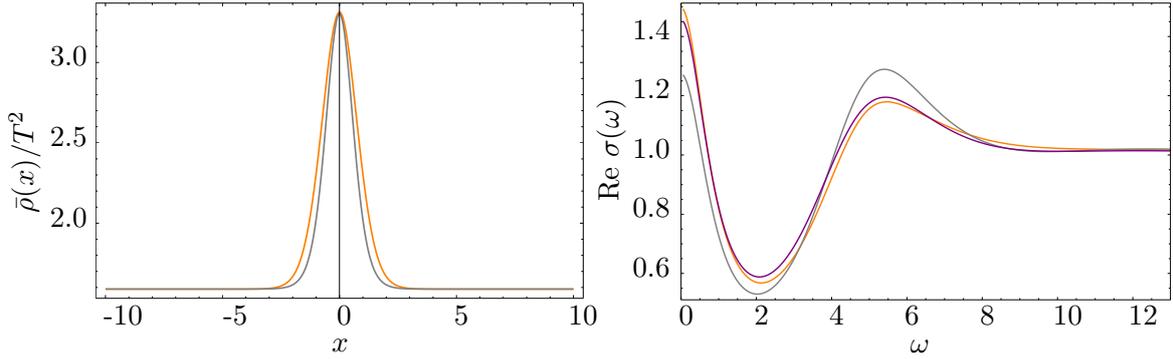%

\caption{On the left panel we plot the charge density (orange line) for a setup with $M=3$ and $\mu=2$. The
gray line corresponds to the charge density of the equivalent homogeneous system at each point $x$,
namely a homogeneous system with the same mass $m(x)$ as our setup at that point. On the right we plot the
same conductivities as in the leftmost plot of fig. \ref{fig:sigmaACyvshomog} (orange and gray lines)
together with the conductivity obtained for an homogeneous system
with $\mu = 3$ and $m=2.095$ (purple line). This last system has the same value of the charge density as the inhomogeneous
setup at the point of interest ($x=0.749$).}
\label{fig:rhoeffect}
\end{figure}

The spatial gradients due to the interface affect the conductivity $\sigma_y(\omega,x)$
in two ways.
The most important effect occurs at the level of the background fields $\chi$ and $\phi$:
the nonzero spatial gradients of these fields result in a value of the charge density $\rho(x)$, which
near the interface is higher than that of a homogeneous system with the same values of mass $m(x)$ and
chemical potential $\mu$. This is shown in the left panel of figure \ref{fig:rhoeffect} 
where we compare both charge densities, and see that indeed, around the interface, the charge density of the
inhomogeneous case is always larger. Consequently, one expects $\sigma_y(x)$ and $\sigma_y^{\rm h}(x)$
to differ, and in particular, the DC value of $\sigma_y(x)$ to be higher than that of $\sigma_y^{\rm h}$.
This is in agreement with what we see in fig. \ref{fig:sigmaACyvshomog}.
Note that this effect, due to the the enhancement of the charge density around the interface, 
would be there even if the form of the equations of motion for the fluctuations 
were not changed with respect to the homogeneous case.
The second effect 
occurs at the level of the equations of motion of the fluctuations. 
Notice that while the fluctuation $a_x$ is coupled to those  of the embedding and the charge density,
$c$ and $a_t$, the field $a_y$ decouples from any other fluctuation, as it happens in the homogeneous case.
However, the DBI action does couple $a_y$ to the spatial derivatives of the background functions,
$\dot\phi$ and $\dot\chi$, as can be seen in \eqref{eq:sdbi2}. Hence, there are new terms
in the equation of motion for $a_y$ that are not present in the homogeneous case, and one expects these
terms to affect the conductivity.
To try and gauge the relevance of these two effects, on the right panel of figure \ref{fig:rhoeffect}
we compare the conductivity $\sigma_y(\omega)$, computed at a point $x$ (orange line),
with the one that results for a system with the same value of the chemical potential and the
charge density as our system at that point $x$ (purple line).
Although the purple line does not exactly overlap with the orange one, it is much closer to it than the gray
line (which as in fig. \ref{fig:sigmaACyvshomog} stands for $\sigma_y^{\rm h}$).
%
This confirms that, as expected, the enhancement of the
charge density due to the spatial inhomogeneities is the dominant effect of the interface on $\sigma_y$.

\begin{figure}
 \centering
 \def\svgwidth{\columnwidth}
\executeiffilenewer{sigmasinterface.svg}{sigmasinterface.pdf}%
{inkscape -z -D --file=sigmasinterface.svg %
--export-pdf=sigmasinterface.pdf --export-latex}%
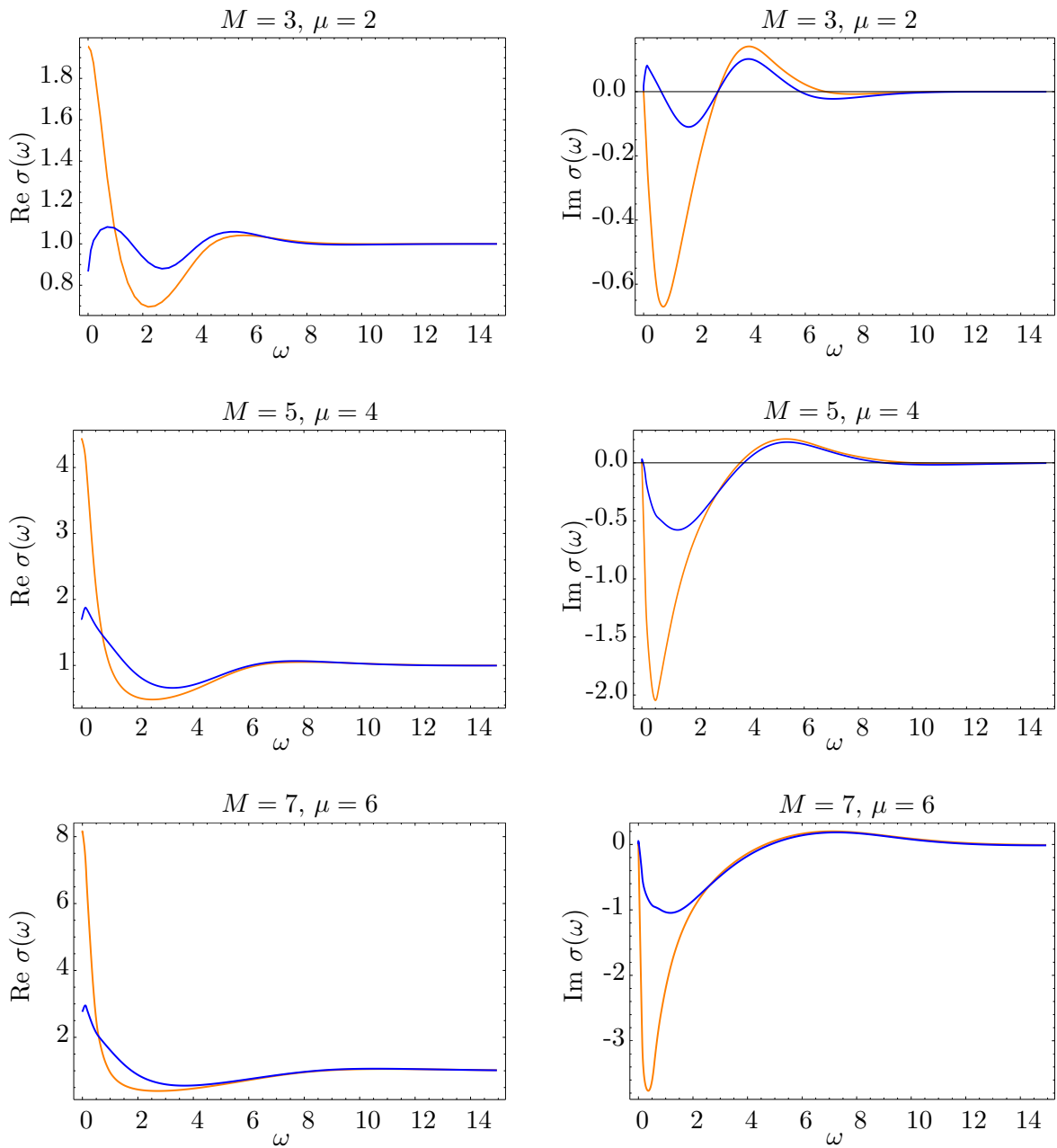%

 \caption{Conductivities at the interface. Plots of the conductivity $\sigma_x(x=0,\omega)$ (blue) and 
$\sigma_y(x=0,\omega)$ (orange) for different values of the background parameters $M$ and $\mu$.
The real parts are shown on the left and the imaginary parts on the right.}
\label{fig:sigmaACxvsy}
\end{figure}

We now focus on the behavior of the conductivities at the interface. In figure \ref{fig:sigmaACxvsy} we plot
both $\sigma_x$ and $\sigma_y$ at the interface for three different values of the background parameters $M$
and $\mu$. We have chosen the  pairs of values $(M,\mu)$ so that they correspond to systems where the 
charge density at the edges is kept low (namely configurations at the edge of the  area accessible to BH 
embeddings in fig. \ref{fig:BHphase}). At the interface the embedding becomes massless,  thus the 
configurations with higher values of $\mu=\bar\mu/T$ correspond to lower temperatures and higher values 
of the charge density. 

By looking at the plots of the real part of the conductivities at the interface, presented on the right 
panels of figure \ref{fig:sigmaACxvsy}, we observe one of the main features of our construction: 
the DC conductivity along the interface ($\sigma_y^{\rm DC}$) is considerably enhanced with respect to 
that in the direction perpendicular to it
($\sigma_x^{\rm DC}$).
This is a direct consequence of the spatial distribution of the charge density in our system (see fig. \ref{fig:chargedensity} for an example of $\rho(x)$). As we have seen in fig. \ref{fig:sigmaACyvshomog}, the conductivity $\sigma_y(\omega)$ is basically determined by the value of the charge density at the point of interest,  in this case $x=0$. However, this is not the case for the conductivity in the $x$ direction $\sigma_x(\omega)$. 
As we discuss below, when focusing on $\sigma_x^{\rm DC}$, the DC conductivity along the $x$ direction,
which must be independent of $x$ due to current conservation, is basically determined by the charge density at the edges of the system, which is much lower than that at the interface. 
Therefore $\sigma_x^{\rm DC}$ is suppressed with respect to $\sigma_y(\omega)$. This suppression is maximal
for embeddings such that the charge density at the edges is arbitrarily small. Nevertheless, 
$\sigma_x^{\rm DC}$  never vanishes completely, since there is always a contribution from the thermally produced 
pairs of charge carriers \cite{Karch:2007pd}.

\subsubsection{DC conductivity}
In the following we focus on the DC conductivity along the direction orthogonal to the interface, namely
$\sigma_x^{\rm DC}$. As is obvious from the action of the fluctuations, $\sigma_x$ is more sensitive to the 
effects of translational symmetry breaking introduced by our inhomogeneous embedding. 
In addition, as  can be seen from the current conservation equation \eqref{eq:currentcons}, 
for a setup like ours, in which the charge density does not vary with time,
$\sigma_x^{\rm DC}$ is a constant. Moreover, as we have described in section \ref{ssec:dcsigma}, 
we can compute $\sigma_x^{\rm DC}$ from 
the behavior of the background functions at the horizon without having to solve the fluctuation equations in 
the $\omega\to0$ limit. Note that eq. \eqref{eq:sigmadexpansion} is 
particularly well suited to numerical 
evaluation for this purpose\footnote{In fact, \eqref{eq:sigmadexpansion} is the expression we evaluate 
numerically to read the value of $\sigma_x^{\rm DC}$. A field redefinition of the form 
$\phi \rightarrow (1-z)^2 \tilde\phi$ eliminates the need to evaluate a term containing a second derivative 
like $a^{(2)}(x)$.}.

\begin{figure}
 \centering
 \def\svgwidth{\columnwidth}
\executeiffilenewer{comparesigmadcS.svg}{comparesigmadcS.pdf}%
{inkscape -z -D --file=comparesigmadcS.svg %
--export-pdf=comparesigmadcS.pdf --export-latex}%
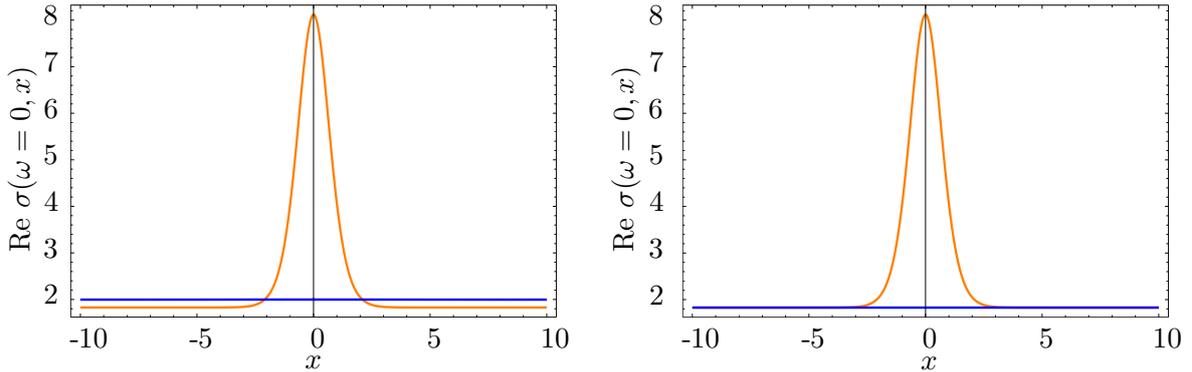%

        \caption{Comparison of $\sigma_x^{DC}(x)$ (blue) and $\sigma_y^{DC}(x)$ 
(orange) for a setup with $\mu=4$ and $M=5.3$. 
On the left panel we plot the results for a short system, for which we set $L=10$. The right panel corresponds to a
long system where $L = 100$. In both cases we have set $a=3$ (remember that $a$ determines the width of the interface via eq. \eqref{eq:massprofile}).}
\label{fig:comparesigmadcS}
\end{figure}

We start by comparing the DC conductivities $\sigma_x^{\rm DC}$ and $\sigma_y^{\rm DC}$. They are plotted in 
figure \ref{fig:comparesigmadcS} for the two kinds of systems introduced in section \ref{ssec:numerics}.
Let us first describe what we expect for $\sigma_y^{\rm DC}$, and then
discuss $\sigma_x^{\rm DC}$ in detail.

The DC conductivity along the direction parallel to the interface, $\sigma_y^{\rm DC}(x)$,
is read from the $\omega\to0$ limit of the AC conductivity $\sigma_y(\omega,x)$.
As shown in fig. \ref{fig:sigmaACyvshomog},
up to a constant \cite{Karch:2007pd} and to some small effects sourced
by the spatial gradients of the background, $\sigma_y^{\rm DC}(x)$ is determined  by the value of the
charge density at each point $x$. Hence, it is expected to peak at the interface, where the charge 
density is maximal, and to asymptote to a nonzero baseline value towards the edges.

In section \ref{ssec:dcsigma} we discussed  how to compute $\sigma_x^{\rm DC}$ in terms of the horizon 
data. Subsequently, in section \ref{ssec:numerics} we defined two different kinds of systems corresponding 
to different boundary conditions for the fluctuations at the edges. As we now show, these result in slightly
different behaviors of $\sigma_x^{\rm DC}$.  

\subsubsection*{Long Systems}
For these systems the effects of the interface fade away towards the edges. Consequently, in eq. \eqref{eq:sigdcdef1} 
we can use the boundary condition \eqref{eq:aydamp}, \emph{i.e.} $\partial_x p(\pm L)=\partial_x a_t(0,\pm L)=0$ to get
\be
\sigma_x^{\rm DC}={\cal F}(z=1,x=\pm L)\,,
\label{eq:sigDCinf}
\ee
which is the DC conductivity of a system without an interface, since ${\cal F}$ at the edges agrees with that of a background with an homogeneous embedding. Notice that this is to be expected; assuming that the effects of the interface do not reach the edges amounts to having a system where the width of the interface is negligible with respect to the total length.  
Therefore we expect $\sigma_x^{\rm DC}$ to be the same as $\sigma_y^{\rm DC}(x=\pm L)$, namely
$\sigma_y^{\rm DC}$ at the edges.

On the right panel of figure \ref{fig:comparesigmadcS} we plot  $\sigma_x^{\rm DC}$ and $\sigma_y^{\rm DC}$ for a long system, and we observe that they overlap away from the interface.

\subsubsection*{Short Systems}
In section \ref{ssec:numerics} we also considered the case of a system
where the effects of the interface reach the boundary by allowing the
derivatives of the fluctuations to be nonzero at the edges. In that
case, it is still possible to
compute $\sigma_x^{\rm DC}$ by means of the integral \eqref{eq:sigdcdef2}.

On the left panel of figure \ref{fig:comparesigmadcS} we plot  $\sigma_x^{\rm DC}$ and 
$\sigma_y^{\rm DC}$ for a short system. We see that $\sigma_x^{\rm DC}$ is slightly 
larger than $\sigma_y^{\rm DC}$ at the edges. The interface is now introducing a sizeable 
region where the charge density is augmented, producing a net enhancement of the conductivity 
$\sigma_x^{\rm DC}$
\footnote{In order to roughly estimate the DC conductivity $\sigma_x^{\rm DC}$, we may think of the system as 
made of two regions: one region of length $\epsilon$ and conductivity $\sigma_{\rm DC}^{\rm int}$, corresponding to the interface; and another region of length $2L-\epsilon$ and conductivity
$\sigma_{\rm DC}^0<\sigma_{\rm DC}^{\rm int}$, corresponding to the system away from the interface.
One can then write
$
\frac{2L}{\sigma_x^{\rm DC}}=\frac{2L-\epsilon}{\sigma_{\rm DC}^0}+\frac{\epsilon}{\sigma_{\rm DC}^{\rm int}}
$. Hence, when $\epsilon\ll L$, $\sigma_x^{\rm DC}=\sigma_{\rm DC}^0$; instead for $\epsilon\lesssim L$, 
we have $\sigma_x^{\rm DC}\gtrsim\sigma_{\rm DC}^0$. We point out that this analogy also works for non-symmetric systems, having $m(x)$ interpolating between different masses at both sides of the interface.}.

\begin{figure}
\centering
\def\svgwidth{0.6\columnwidth}
\executeiffilenewer{object.svg}{object.pdf}%
{inkscape -z -D --file=object.svg %
--export-pdf=object.pdf --export-latex}%
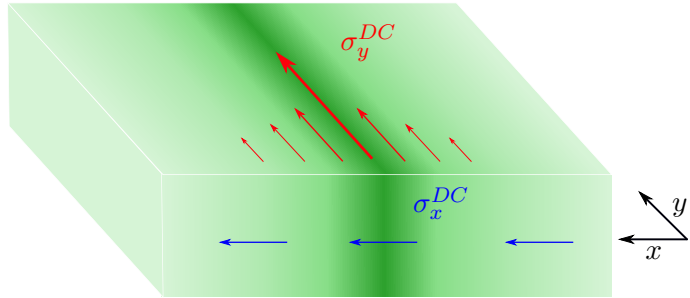%

\caption{Schematic illustration of our setup. The system is two-dimensional; the third (vertical)
dimension has been added for illustrative purposes. The intensity of green encodes the $x$-dependence 
of the charge density (see fig. \ref{fig:chargedensity}), with darker
green standing for larger charge 
density. The red arrows represent the value of $\sigma_y^{\rm DC}$, which varies along the system, 
and  the blue arrows denote the value  of $\sigma_x^{\rm DC}$, which is constant.}
\label{fig:object}
\end{figure}

In figure \ref{fig:object} we illustratively summarize the behavior of the DC conductivities in our system. As is clear from the illustration, $\sigma_y^{\rm DC}(x)$ roughly follows the charge density, which varies along $x$ and peaks at the interface, while $\sigma_x^{\rm DC}$ is constant, its value mainly determined by the charge density away from the interface.

\begin{figure}[b]
\centering
\def\svgwidth{\columnwidth}
\executeiffilenewer{tempsnew.svg}{tempsnew.pdf}%
{inkscape -z -D --file=tempsnew.svg %
--export-pdf=tempsnew.pdf --export-latex}%
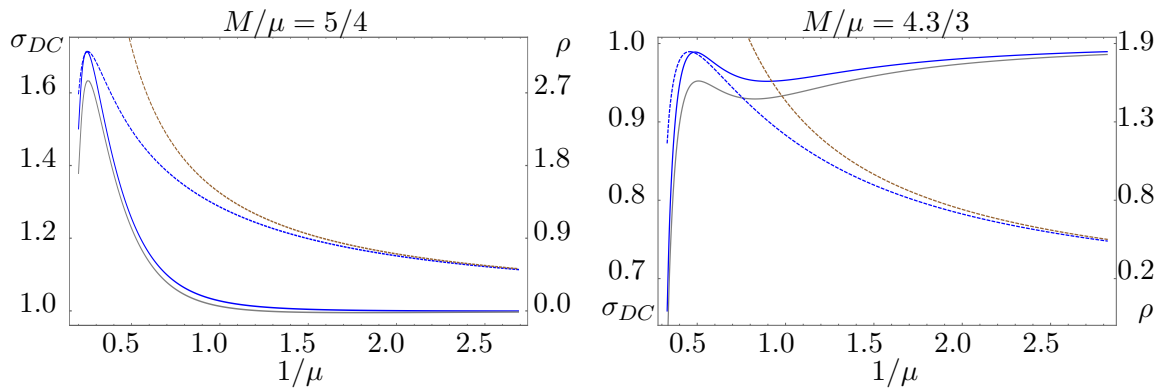%

\caption{DC conductivity  versus $1/\mu = T/\bar\mu$. The solid blue line corresponds to $\sigma_x^{\rm DC}$,
while the solid gray line represents $\sigma_{\rm DC}^0$, which corresponds to a system with no interface. For guidance we also plot the values of the charge density at the edges (blue dashed lines) and at the interface (gray dashed lines). The left panel corresponds to a setup with $\bar\mu/M_q =4/5$. The right panel is for $\bar\mu/M_q =3/4.3$.
The left vertical axes refer to the conductivity plots whereas the right ones show the scale for the charge density.}
\label{fig:sigmadctemp}
\end{figure}

Finally, in figure \ref{fig:sigmadctemp} we study the evolution of $\sigma_x^{\rm DC}$ as a function of 
$1/\mu = T/\bar\mu$, at fixed $M/\mu=M_q/\bar\mu$. We perform the analysis for a short system with 
$L=10$, and in order to study the effect of the interface, we compare $\sigma_x^{\rm DC}$ to the 
conductivity of an equivalent homogeneous system $\sigma_{\rm DC}^0$. This is the DC conductivity 
for an homogeneous system with the same mass $M$ and chemical potential $\mu$ as our setup at its edges.
Notice that for the values of $M/\mu$ considered, there is a minimum value of $1/\mu$ that can be 
reached by our embeddings (see fig. \ref{fig:phasediag2}). Both conductivities grow as we lower the 
temperature ($1/\mu$) until they reach a maximum, and then decrease rapidly. As is clear from the plots, 
the behavior of the conductivity follows closely that of the charge density at the edges 
(blue dashed line). Moreover, $\sigma_x^{\rm DC}$ is always slightly larger than the homogeneous 
counterpart $\sigma_{\rm DC}^0$ as expected from the length of the system. Also note that, with 
respect to an homogeneous system, the presence of the interface has two competing effects on 
the DC conductivity. On the one hand, the interface is a region of small size where the charge density 
is much larger than in the homogeneous system towards the edges. This fact, as we have seen when 
discussing the plots in fig. \ref{fig:comparesigmadcS}, should result in an enhancement of 
$\sigma_x^{\rm DC}$. On the other hand, the presence of the interface gives rise to inhomogeneities 
that, on general grounds, should impede the conductivity \cite{Ryu:2011vq}. 
However, in view of our results the enhancement of the charge density is strong enough to overcome 
other effects of the inhomogeneous embedding.

\section{Conclusions 
}
\label{sec:conclusion}
In this work we constructed a holographic system that realizes a one-dimensional charged interface.
We have considered a D3/D5 intersection in the limit 
where the number of D5-branes is much smaller than that of D3-branes. The gravity dual consists of a probe 
D5-brane in the background generated by the D3-branes.
We are interested in systems at fine temperature, which is why we have embedded the probe D5-brane in a black 
D3-brane geometry. Furthermore, in order to have a finite charge density of the fundamental degrees of freedom 
sourced by the D5-branes we switched on the temporal component of the worldvolume gauge field of the probe.
Our configurations are massive embeddings for which the asymptotic distance between the D5-branes and the 
D3-branes is non-vanishing.
The D5-brane is extended along two spatial Minkowski directions, giving rise to a (2+1)-dimensional defect. 
To realize the interface, we chose the mass parameter to depend on one of those Minkowski directions, which 
we denote by $x$, with a profile that interpolates between two homogeneous embeddings corresponding to a 
positive and negative mass respectively.
Due to technical limitations arising from the numerics, our embeddings are
chosen to be of black hole type for every value of $x$, thus the induced charge density is nonzero along the 
entire system. Nevertheless, the charge density peaks at the interface, where the mass vanishes, and away from 
it can be made arbitrarily small by appropriately tuning the embedding.
To sum up, the embedding is described by two fields: the embedding scalar $\chi$ and the temporal component of the worldvolume gauge field $\phi$. They both depend on the radial coordinate $z$ and on the spatial direction $x$. The dynamics of the probe brane is governed by the DBI action and the equations of motion are a set of two coupled non-linear PDEs, for $\chi$ and $\phi$. In order to construct our holographic interface we have solved those PDEs numerically.

For the configurations obtained in this way, we checked  that indeed the charge density peaks at the interface. 
For the examples provided, the value of the charge density at the interface is about five times larger than at 
the edges. Moreover, both at the interface and away from it, we have studied the scaling of the charge density 
with the chemical potential, finding that it agrees with the result for the homogeneous intersection, 
$\rho\propto\mu^2$ \cite{Nogueira:2011sx}.

After constructing the backgrounds realizing the holographic
interface, we  
proceeded with the computation of the electrical conductivities of the system. For this purpose, we studied the fluctuations of the worldvolume gauge field, which in some cases couple among themselves and with that of the scalar describing the embedding. We now summarize our main results.

We have computed the AC and DC conductivities both in the direction perpendicular to the interface
($\sigma_x$) and in that parallel to it ($\sigma_y$).
As for the AC conductivities, away from the interface $\sigma_x$ and $\sigma_y$ do not present substantial
differences and almost agree with the AC conductivity of an homogeneous system without interface.
In particular, we observe the peaks due to mesonic excitations known
from brane embeddings constant in $x$. 
These peaks disappear at the interface where the charge carriers become massless. Moreover, at the
interface $\sigma_x$ and $\sigma_y$ present differences at low frequencies. 
Notice that low frequencies correspond to probing the system at long ranges, and
for large length scales the setup looks very different in the $x$ and in the $y$ direction.

Next, we studied the effects of the inhomogeneous background on the conductivity in the direction 
parallel to the interface, $\sigma_y$. 
Despite the background being homogeneous along the $y$ direction, due to the non-linearities of the DBI 
action, a current along $y$ is sensitive to the gradients along $x$ of the background fields $\chi$ and $\phi$.
These gradients are larger in the vicinity of the interface, and therefore
have  more significant effects upon the conductivity $\sigma_y$ in
this region.
We have checked that the main effect of these gradients on the conductivity $\sigma_y$
is due to  an enhancement of the charge density with respect to what would be the equivalent homogeneous system.
In particular, we found that close to the interface the inhomogeneous background
causes a transfer of spectral weight from mid to low frequencies, resulting in an enhancement 
of the DC conductivity $\sigma_y^{\rm DC}$.

We have paid special attention to the computation of the DC conductivity in the direction perpendicular to the interface,  $\sigma_x^{\rm DC}$. 
For a system like the one at hand, where the charge density is inhomogeneous only in the $x$ direction, current conservation implies that $\sigma_x^{\rm DC}$ is constant.
Moreover, following \cite{Iqbal:2008by,Ryu:2011vq}, we were able to compute $\sigma_x^{\rm DC}$
in terms of the background fields evaluated at the horizon. 
Two of the most relevant results of this work are related to the conductivity $\sigma_x^{\rm DC}$.

An important feature of our system is the fact that $\sigma_x^{\rm DC}$ is basically determined by the homogeneous system away from the interface, where the charge density is very low.
On the other hand, $\sigma_y^{\rm DC}$ varies with $x$ and is roughly proportional to the charge density.
Consequently, at the interface the conductivity along it, $\sigma_y^{\rm DC}$, is considerably larger than the conductivity
in the orthogonal direction, $\sigma_x^{\rm DC}$. For some of our numerical simulations,
$\sigma_y^{\rm DC}$ at the interface is up to $4\times\sigma_x^{\rm DC}$.

Our  analysis of $\sigma_x^{\rm DC}$ provides information on
the effects of the breaking of translational 
symmetry caused by our interface. In particular, we showed that when the size of the 
inhomogeneous region is not negligible with respect to the total size of the system,
$\sigma_x^{\rm DC}$ is sensitive to the inhomogeneities. 
Interestingly, two competing effects are expected to be at work in this scenario: the increase of charge 
density at the interface is expected to cause an increase of $\sigma_x^{\rm DC}$, while the non-vanishing 
gradient of the charge density along $x$ ought to suppress the conductivity \cite{Ryu:2011vq}. 
In the light of our results, we see the effect of charge density localization being clearly dominant 
for our setup. Additionally, we studied the evolution of $\sigma_x^{\rm DC}$ with the temperature, 
and found it to be always larger than its homogeneous counterpart, with the increment becoming larger at 
lower temperatures.

To conclude, let us comment on some possible applications of the holographic interface studied in this work.
The D3/D5 intersection with a kink profile for the embedding was
proposed in \cite{Karch:2010mn} as the holographic realization  of
Quantum Spin Hall (QSH) insulators. A key feature of these systems is
a non-vanishing mixed Chern-Simons term of the form $A^R\wedge
dA$. The gauge field $A^R$ corresponds to a $U(1)$ $R$-symmetry, which
in this condensed matter systems is associated to the $U(1)$ spin
global symmetry, i.e.~the $z$ component of the spin of the electrons. In the holographic dual, the $U(1)_R$ corresponds to a shift symmetry of the internal $S^5$, and the gauge field $A^R$ appears as a fluctuation of the RR four-form $C_4$
\cite{Aharony:1999rz}. A non-trivial WZ term of the form $A^R\wedge dA$ is generated when fluctuations of the $C_4$ are considered.
It would therefore be interesting to extend the analysis of this work
to the case in which a topological term as described above is present. 
This would allow for a study of the QSH conductivity in a (2+1)-dimensional topological insulator.

Another interesting continuation of our work would be the extension of our analysis to the case of the holographic (3+1)-dimensional topological insulators constructed in
\cite{HoyosBadajoz:2010ac} by means of a D3/D7 intersection. That system consists of a D7-brane probe with an inhomogeneous embedding as the one studied here. In that case a nonzero WZ term is generated at the interface, sourcing a finite Hall conductivity.
The relevant D7-brane embeddings at finite temperature and charge
density were constructed in \cite{Rozali:2012gf}, and the next natural step
would be the repetition of the analysis of the conductivity presented here for
those D7-brane 
embeddings. A crucial difference would arise from the nonzero Hall
conductivity around the interface. Since the embeddings
at finite charge density are always of black hole type, this model
would describe the electric transport at the transition between different quantum Hall states, see \cite{Alanen:2009cn}.

A further continuation of this work would consist in trying to ascertain the existence of purely fermionic
massless degrees of freedom at the interface. For a embedding with a step-like profile,
at the interface the D5-brane shares only one spatial direction with the D3-branes, and therefore the
intersection becomes of the D3/D5' type.
%
%
For this intersection, with six mixed boundary conditions on the worldsheet ($\# ND=6$), the only fundamental 
massless degrees of freedom in the spectrum are fermionic.
Hence it would be interesting to study the Green's functions of different (gauge invariant) operators at such
an interface, and look for signatures of the existence of a Fermi surface.

%

Finally, as proposed in \cite{Ryu:2011vq}, setups involving
inhomogeneous brane intersections may be used
to study the effects of disorder on strongly coupled systems at finite
charge density. An example for this is provided by intersections where the chemical potential is given by an 
inhomogeneous profile with random fluctuations around a baseline value, as in \cite{Arean:2013mta}. 
This is a direction of work we are actively pursuing and we expect to report on our results in the future.

\acknowledgments
It is a pleasure to thank M. Ammon, O.~Bergman, A. Donos, M. Jarvinen, A. Karch, 
G.~Lifshytz, C. Pantelidou, M. P\'erez-Victoria, M. Rozali, I. Salazar, J. Tarr\'\i o, M. Yunpu and H. 
Zeller for useful discussions. D.A.~thanks the FRont Of pro-Galician
Scientists for unconditional support. The work of D.~Are\'an is
supported by the German-Israeli Foundation (GIF), grant 1156.
The work of J.M. Lizana is supported by MINECO, via the FPI program, 
and projects FPA2010-17915 and FPA2013-47836-C3-2-P and by the 
Junta de Andaluc\'ia projects FQM 101 and FQM 6552.

\appendix
\section{Background equations of motion}
\label{app:eoms}
In this appendix we write down the equations of motion of the background functions, namely
those for the functions $\chi(z,x)$ and $\phi(z,x)$ describing the embedding of the probe D5-brane.
They follow from the action \eqref{eq:action}, and take the form
\begin{align}
 &\eta_1 (\partial_z^2 \phi)+\eta_2 (\partial_x^2 \phi)
 +\eta_3 (\partial_z\partial_x \phi)+\eta_4 (\partial_z \phi)^3
 +\eta_5 (\partial_z\chi \partial_x \chi \partial_x\phi)+\eta_6 (\partial_z \phi)=0\,,\\
 &\tau_1 (\partial_z^2 \chi)+\tau_2 (\partial_x^2 \chi)+\tau_3 (\partial_z\partial_x \chi)
 +\tau_4 (\partial_z \chi)^3+\tau_5 (\partial_z\chi)^2+\tau_6 (\partial_z \chi)
 +\tau_7(\partial_x\chi)^2\nonumber\\
 &\,+\tau_8(\partial_x\chi)+\tau_9 \chi=0,
 \label{eq:phieom}
\end{align}
where the coefficients $\eta_i$, ($i=1\dots 7$), and $\tau_i$, ($i=1\dots 9$),
are given by the following functions of $z$ and $x$,
\begin{align}
 \eta_1(z,x)= &\, 2h\left[h z^4 (1-\chi^2)\dot\phi^2-f^2(h(1-\chi^2)+z^2\dot\chi^2) \right]\,,\nn\\
 \eta_2(z,x)= &\, 2h\left[h z^4 (1-\chi^2)\phi'^2-f^2(1-\chi^2+z^2\chi'^2) \right]\,,\nn\\
 \eta_3(z,x)= &\, 4 h z^2 \left[f^2 \chi' \dot\chi-h z^2 (1-\chi^2)\phi'\dot\phi\right]\,,\nn\\
 \eta_4(z,x)= &\, h z^3 (z h'-2h) \left[ 2 h(1-\chi^2) +z^2 \dot\chi^2\right]\,,\nn\\
 \eta_5(z,x)= &\, 4 f h z (2f - z f')+2 h z^5 \phi'^2(2h-zh')\,,\nn\\
 \eta_6(z,x)= &\, h z^3 \dot\phi^2 \left[-h\left(4(1-\chi^2)+2 z^2 \chi'^2\right)+z h' \left( 3(1-\chi^2)+z^2 \chi'^2\right)\right]\nn\\
  &-2 f h f' \left[-z^2 \dot\chi^2 - h \left(1-\chi^2-z^2 \chi'^2 \right)\right]+f^2\left[-2 z^2 \dot\chi^2 h'+6 h^2 z \chi'^2\right.\nn\\
  &+ \left. h \left(-2 z \dot\chi^2-h'(3(1-\chi^2)+z^2 \chi'^2)\right)\right]\nn\\
 \end{align}
 \begin{align}
  \tau_1(z,x)=&\,2 h z^2\left[h z^4 (1-\chi^2) \dot\phi^2-f^2(h(1-\chi^2)+z^2 \dot\chi^2)\right], \nn\\
  \tau_2(z,x)=&\,2 h z^2\left[h z^4(1-\chi^2)\phi'^2-f^2(1-\chi^2+z^2\chi'^2)\right], \nn\\
  \tau_3(z,x)=&\,4 h z^4 \left[f^2 \chi' \dot\chi-h z^2 \phi' \dot\phi (1-\chi^2) \right], \nn\\
  \tau_4(z,x)=&\,h z^3\left[6 f^2 h - 2 f h z f'-2 h z^4\dot\phi^2-z h' (f^2-z^4 \dot\phi^2)\right] \nn\\
  \tau_5(z,x)=&\,6 h^2 z^2 \chi (-f^2+z^4 \dot\phi^2)+2 h z^7 \phi' (2 h -z h')\dot\phi\dot\chi, \nn\\
  \tau_6(z,x)=&\,z \Big\{-2 f h z \left(h(1-\chi^2)+z^2 \dot\chi^2\right)f'+f^2\left[4 h^2(1-\chi^2)-2 z^3 \dot\chi^2 h'\right.\nn\\ 
   &+h z \left.\left(6 z \dot\chi^2 -(1-\chi^2)h'\right)\right]+ h z^4 \left[ -2h\phi'\left(6z\chi \dot\phi \dot\chi - 2 h \chi^2 \phi' + (2h+z^2 \dot\chi^2)\right)\right.\nn\\ 
   &+z h' \left.\left.\left((1-\chi^2)\dot\phi^2+\left(2 h (1-\chi^2) + z^2 \dot\chi^2\right)\phi'^2\right)\right]\right\}\nn \\
  \tau_7(z,x)=&\,6 h z^2 \chi \left(h z^4 \phi'^2-f^2\right)\, ,\nn\\
  \tau_8(z,x)=&\,-2 h z^5 (1-\chi^2) (2 h - z h')\phi' \dot\phi \,,\nn\\
  \tau_9(z,x)=&\,4 h ^2 \left[z^4\left( \dot\phi^2+h \phi'^2\right)-f^2\right]\,,
\end{align}
where primes stand for derivatives with respect to $z$, and dots for derivatives with respect to $x$.

\section{Quadratic action for the fluctuations}
\label{app:action}
In this appendix we present the action of the fluctuations considered in section \ref{ssec:flucts},
and which allowed us to compute the conductivity of the setup.
The action results from expanding the DBI action up to second order in the fluctuations \eqref{eq:flucdef},
and can be written as
\be
S^{(2)}=-N_f T_{D5}\,L^6
\int dt\,d^2x\,dz\,d\Omega_2\, {\cal L}^{(2)}\,,
\label{eq:action2}
\ee  
with
\begin{equation}
 \begin{split}
{\cal L}^{(2)}=& -\Delta \Bigg\{ \Bigg[ c \Upsilon \chi-z^2 (1-\chi^2)(1-\chi^2)(a_t'
-i \omega a_z)h^2 z^2 \dot\phi-f^2 \chi' \dot c+h\bigg(z^4\dot\phi \dot c (\dot\phi \chi'-\dot\chi \phi')\\
& -c' \left(\dot\chi(f^2-z^4 \phi'^2)+z^4 \dot\phi \phi' \chi'\right)+z^4 \chi'(\dot\phi \chi'
-\dot\chi \phi')(a_t'-i\omega a_z)\\
& +\left(z^4\dot\phi \dot\chi \chi'-z^2(1-\chi^2+z^2 \dot\chi^2)\right)(i\omega a_x -\dot a_t)\bigg)\Bigg)\Bigg]^2\\
& -\frac{\Sigma}{h(1-\chi^2)}\Bigg[\Omega h c^2 - 4 h z^2 \chi c \dot c \left(h z^4 \chi' \dot\phi^2-f^2 \chi'
-h z^4\dot\phi \dot\chi \phi'\right)\\
&- h\left(z^4 \dot\phi \phi' \chi' + \dot\chi(f^2-z^4 \phi'^2)\right)c'
+h z^2 \left(\dot\phi\left(2h(1-\chi^2)+z^2 \chi'^2\right)
-z^2 \dot\chi\phi'\chi' \right)(a_t'-i\omega a_z)\\
&+ \left(h z^4 \dot\phi \dot\chi \chi'-h z^2 (2-2 \chi^2+z^2 \dot\chi^2)\right)(i\omega a_x -\dot a_t)\\
&- z^2 (1-\chi^2)\Bigg(h^3 \left( 2 z^2 i \omega (1-\chi^2)a_t' a_z-z^2 (1-\chi^2)a_t'^2
+\omega^2\left( c^2+z^2(1-\chi^2)\right) a_z^2\right)\\
&+ f^2 z^2 \bigg(z^2 \chi'^2 a_y'^2-2 z^2 \dot\chi \chi' a_y' \dot a_y 
+ (1-\chi^2+z^2 \dot\chi^2)\dot a_y^2\bigg)+h \bigg( f^2 z^2 (1-\chi^2) a_x'^2+f^2 \dot c^2\\
&+ z^2 (1-\chi^2)(f^2-z^4 \phi'^2) a_y'^2+z^4 \omega^2 \chi'^2 a_y^2
+2 z^6 (1-\chi^2)\dot\phi \phi' a_y' \dot a_y-z^6 \dot\phi^2 \dot a_y^2+z^6 \chi^2 \dot\phi^2 \dot a_y^2\\
&- 2 f^2 z^2 (1-\chi^2) a_x' a_z'+f^2 z^2 a_z'^2-f^2 z^2 \chi^2 a_z'^2 \bigg)
-h^2\bigg(2 z^4 \dot\phi \phi' c' \dot c-(f^2-z^4 \phi'^2)c'^2\\
&- \phi'\chi'(a_t'-i\omega a_z)+(\dot\phi \chi'-2\dot\chi \phi')(i \omega a_x-\dot a_t)\\
& +z^2 \Big(z^2 \dot\phi^2 \dot c^2+2 z^2 \dot\chi \phi' (i \omega c a_x'-\dot c a_t' + i \omega \dot c a_z )-(1-\chi^2+z^2 \dot\chi^2) \omega^2 (a_x^2+a_y^2)\\
&+ 2 z^2  \dot\phi \chi' (2 \dot c a_t'- i \omega \dot c a_z-i \omega c a_x')
+z^2 \chi'^2 (a_t'^2-2 i \omega a_t' a_z-\omega^2 a_z^2)\\
&+ 2 z^2 \dot\phi \dot\chi \left((i \omega a_x -\dot a_t)\dot c \right)
-2 z^2 \dot\chi \chi' \left((i \omega a_z-a_t' )(i \omega a_x - \dot a_t)\right)\\
&+(1-\chi^2+z^2 \dot\chi^2)(\dot a_t^2-2 i \omega a_x \dot a_t)- 2 z^2  \left( \dot\chi \phi'+\dot\phi \chi'\right) i \omega c a_z \Big)\bigg)\Bigg]\Bigg\},\\
\end{split}
\label{eq:sdbi2}
\end{equation}
where
\begin{equation}
 \Delta=\frac{1}{2 \,z^{16}\, {\cal L}^{(0)\,3}},
\end{equation}
${\cal L}^{(0)}$ being the Lagrangian of the zeroth order DBI action, \eqref{eq:action}
and the functions $\Upsilon$, $\Sigma$ and $\Omega$ are defined as
\begin{align}
 \Upsilon &= h z^4 \left[(2-2\chi^2+z^2\, \dot\chi^2)\phi'^2-2 z^2\, \dot\phi\,\dot\chi\,\phi'\,\chi'
  +(2h\,(1-\chi^2)+z^2 \chi'^2)\,\dot\phi^2\right]\nonumber\\
  &\quad -f^2 \left[h\,(2-2\chi^2+z^2 \dot\chi^2)+z^2\, \chi'^2\right],  \\
    \Sigma &= (1-\chi^2)^2 \bigg\{h\, z^4 \left[(1-\chi^2+z^2\, \dot\chi^2)\,\phi'^2
    -2 z^2\, \dot\phi\,\dot\chi\,\phi'\,\chi'+\left(h\,(1-\chi^2)
    +z^2\, \chi'^2\right)\dot\phi^2\right]\nonumber\\
  &\quad -f^2 \left[h\,(1-\chi^2+z^2\, \dot\chi^2)+z^2\, \chi'^2\right]\bigg\},   \\
  \Omega &= \bigg\{h z^4 \left[(2-6\chi^2+z^2\, \dot\chi^2)\,\phi'^2-2 z^2\, \dot\phi\,\dot\chi\,\phi'\,\chi'
  +(2h\,(1-3\chi^2)+z^2\, \chi'^2)\,\dot\phi^2\right]\nonumber\\
  &\quad -f^2 \left[h\,(2-6\chi^2+z^2\, \dot\chi^2)+z^2\, \chi'^2\right]\bigg\}\,.
  \end{align}
Note that we gauge away the radial gauge field component $a_z$ once the equations of motion for the fluctuation fields have been found. 

\bibliography{report}

\end{document}

%% file: phasediag.pdf_tex
\begingroup%
  \makeatletter%
  \providecommand\color[2][]{%
    \errmessage{(Inkscape) Color is used for the text in Inkscape, but the package 'color.sty' is not loaded}%
    \renewcommand\color[2][]{}%
  }%
  \providecommand\transparent[1]{%
    \errmessage{(Inkscape) Transparency is used (non-zero) for the text in Inkscape, but the package 'transparent.sty' is not loaded}%
    \renewcommand\transparent[1]{}%
  }%
  \providecommand\rotatebox[2]{#2}%
  \ifx\svgwidth\undefined%
    \setlength{\unitlength}{527.51669922bp}%
    \ifx\svgscale\undefined%
      \relax%
    \else%
      \setlength{\unitlength}{\unitlength * \real{\svgscale}}%
    \fi%
  \else%
    \setlength{\unitlength}{\svgwidth}%
  \fi%
  \global\let\svgwidth\undefined%
  \global\let\svgscale\undefined%
  \makeatother%
  \begin{picture}(1,0.64292308)%
    \put(0,0){\includegraphics[width=\unitlength]{phasediag.pdf}}%
    \put(0.05216271,0.03401562){\color[rgb]{0,0,0}\makebox(0,0)[lb]{\smash{0.0}}}%
    \put(0.19539405,0.03401562){\color[rgb]{0,0,0}\makebox(0,0)[lb]{\smash{0.5}}}%
    \put(0.33862539,0.03401562){\color[rgb]{0,0,0}\makebox(0,0)[lb]{\smash{1.0}}}%
    \put(0.48185673,0.03401562){\color[rgb]{0,0,0}\makebox(0,0)[lb]{\smash{1.5}}}%
    \put(0.62508806,0.03401562){\color[rgb]{0,0,0}\makebox(0,0)[lb]{\smash{2.0}}}%
    \put(0.76831941,0.03401562){\color[rgb]{0,0,0}\makebox(0,0)[lb]{\smash{2.5}}}%
    \put(0.91155074,0.03401562){\color[rgb]{0,0,0}\makebox(0,0)[lb]{\smash{3.0}}}%
    \put(0.04038223,0.05600544){\color[rgb]{0,0,0}\makebox(0,0)[lb]{\smash{0}}}%
    \put(0.03628151,0.17714059){\color[rgb]{0,0,0}\makebox(0,0)[lb]{\smash{1}}}%
    \put(0.04021238,0.29827573){\color[rgb]{0,0,0}\makebox(0,0)[lb]{\smash{2}}}%
    \put(0.03885356,0.41941088){\color[rgb]{0,0,0}\makebox(0,0)[lb]{\smash{3}}}%
    \put(0.04057635,0.54054602){\color[rgb]{0,0,0}\makebox(0,0)[lb]{\smash{4}}}%
    \put(0.55372437,0.09460553){\color[rgb]{0,0,0}\makebox(0,0)[lb]{\smash{0.00}}}%
    \put(0.64996341,0.09460553){\color[rgb]{0,0,0}\makebox(0,0)[lb]{\smash{0.05}}}%
    \put(0.74620242,0.09460553){\color[rgb]{0,0,0}\makebox(0,0)[lb]{\smash{0.10}}}%
    \put(0.84244145,0.09460553){\color[rgb]{0,0,0}\makebox(0,0)[lb]{\smash{0.15}}}%
    \put(0.93868047,0.09460553){\color[rgb]{0,0,0}\makebox(0,0)[lb]{\smash{0.20}}}%
    \put(0.5211491,0.11962843){\color[rgb]{0,0,0}\makebox(0,0)[lb]{\smash{0.0}}}%
    \put(0.52057282,0.1857162){\color[rgb]{0,0,0}\makebox(0,0)[lb]{\smash{0.5}}}%
    \put(0.51982971,0.25180397){\color[rgb]{0,0,0}\makebox(0,0)[lb]{\smash{1.0}}}%
    \put(0.51925343,0.31789173){\color[rgb]{0,0,0}\makebox(0,0)[lb]{\smash{1.5}}}%
    \put(0.01422645,0.30616959){\color[rgb]{0,0,0}\rotatebox{90}{\makebox(0,0)[lb]{\smash{$M_q/T$}}}}%
    \put(0.4934477,0.00187494){\color[rgb]{0,0,0}\makebox(0,0)[lb]{\smash{$\bar\mu/T$}}}%
    \put(0.4934477,0.00187494){\color[rgb]{0,0,0}\makebox(0,0)[lb]{\smash{$\bar\mu/T$}}}%
  \end{picture}%
\endgroup%

%% file: phasediag2.pdf_tex
\begingroup%
  \makeatletter%
  \providecommand\color[2][]{%
    \errmessage{(Inkscape) Color is used for the text in Inkscape, but the package 'color.sty' is not loaded}%
    \renewcommand\color[2][]{}%
  }%
  \providecommand\transparent[1]{%
    \errmessage{(Inkscape) Transparency is used (non-zero) for the text in Inkscape, but the package 'transparent.sty' is not loaded}%
    \renewcommand\transparent[1]{}%
  }%
  \providecommand\rotatebox[2]{#2}%
  \ifx\svgwidth\undefined%
    \setlength{\unitlength}{585.50244141bp}%
    \ifx\svgscale\undefined%
      \relax%
    \else%
      \setlength{\unitlength}{\unitlength * \real{\svgscale}}%
    \fi%
  \else%
    \setlength{\unitlength}{\svgwidth}%
  \fi%
  \global\let\svgwidth\undefined%
  \global\let\svgscale\undefined%
  \makeatother%
  \begin{picture}(1,0.64017266)%
    \put(0,0){\includegraphics[width=\unitlength]{phasediag2.pdf}}%
    \put(0.08691435,0.59815338){\color[rgb]{0,0,0}\makebox(0,0)[rb]{\smash{2.5}}}%
    \put(0.08691435,0.50008239){\color[rgb]{0,0,0}\makebox(0,0)[rb]{\smash{2.0}}}%
    \put(0.08560938,0.40201278){\color[rgb]{0,0,0}\makebox(0,0)[rb]{\smash{1.5}}}%
    \put(0.08560938,0.30394318){\color[rgb]{0,0,0}\makebox(0,0)[rb]{\smash{1.0}}}%
    \put(0.0867142,0.20587219){\color[rgb]{0,0,0}\makebox(0,0)[rb]{\smash{0.5}}}%
    \put(0.0867142,0.10780258){\color[rgb]{0,0,0}\makebox(0,0)[rb]{\smash{0.0}}}%
    \put(0.12193045,0.03807102){\color[rgb]{0,0,0}\makebox(0,0)[rb]{\smash{0.0}}}%
    \put(0.24746502,0.03807102){\color[rgb]{0,0,0}\makebox(0,0)[rb]{\smash{0.5}}}%
    \put(0.37304252,0.03807102){\color[rgb]{0,0,0}\makebox(0,0)[rb]{\smash{1.0}}}%
    \put(0.4985771,0.03807102){\color[rgb]{0,0,0}\makebox(0,0)[rb]{\smash{1.5}}}%
    \put(0.62405577,0.03807102){\color[rgb]{0,0,0}\makebox(0,0)[rb]{\smash{2.0}}}%
    \put(0.74959031,0.03807102){\color[rgb]{0,0,0}\makebox(0,0)[rb]{\smash{2.5}}}%
    \put(0.87512986,0.03807902){\color[rgb]{0,0,0}\makebox(0,0)[rb]{\smash{3.0}}}%
    \put(1.00066448,0.03807902){\color[rgb]{0,0,0}\makebox(0,0)[rb]{\smash{3.5}}}%
    \put(0.58592758,0.00325842){\color[rgb]{0,0,0}\makebox(0,0)[rb]{\smash{$T/M_q$}}}%
    \put(0.01281752,0.41545806){\color[rgb]{0,0,0}\rotatebox{90}{\makebox(0,0)[rb]{\smash{$\bar\mu/M_q$}}}}%
  \end{picture}%
\endgroup%

%% file: afieldb.pdf_tex
\begingroup%
  \makeatletter%
  \providecommand\color[2][]{%
    \errmessage{(Inkscape) Color is used for the text in Inkscape, but the package 'color.sty' is not loaded}%
    \renewcommand\color[2][]{}%
  }%
  \providecommand\transparent[1]{%
    \errmessage{(Inkscape) Transparency is used (non-zero) for the text in Inkscape, but the package 'transparent.sty' is not loaded}%
    \renewcommand\transparent[1]{}%
  }%
  \providecommand\rotatebox[2]{#2}%
  \ifx\svgwidth\undefined%
    \setlength{\unitlength}{1151.45996094bp}%
    \ifx\svgscale\undefined%
      \relax%
    \else%
      \setlength{\unitlength}{\unitlength * \real{\svgscale}}%
    \fi%
  \else%
    \setlength{\unitlength}{\svgwidth}%
  \fi%
  \global\let\svgwidth\undefined%
  \global\let\svgscale\undefined%
  \makeatother%
  \begin{picture}(1,0.38775251)%
    \put(0,0){\includegraphics[width=\unitlength]{afieldb.pdf}}%
    \put(0.43851246,0.16229379){\color[rgb]{0,0,0}\makebox(0,0)[lb]{\smash{-1}}}%
    \put(0.44443688,0.20857121){\color[rgb]{0,0,0}\makebox(0,0)[lb]{\smash{0}}}%
    \put(0.44643845,0.25938352){\color[rgb]{0,0,0}\makebox(0,0)[lb]{\smash{1}}}%
    \put(0.24934024,0.36257273){\color[rgb]{0,0,0}\makebox(0,0)[lb]{\smash{0.0}}}%
    \put(0.33369273,0.32797391){\color[rgb]{0,0,0}\makebox(0,0)[lb]{\smash{0.5}}}%
    \put(0.43205628,0.28451094){\color[rgb]{0,0,0}\makebox(0,0)[lb]{\smash{1.0}}}%
    \put(0.36294862,0.35706433){\color[rgb]{0,0,0}\makebox(0,0)[lb]{\smash{$z$}}}%
    \put(0.16633713,0.03732814){\color[rgb]{0,0,0}\makebox(0,0)[lb]{\smash{-10}}}%
    \put(0.24092747,0.06477843){\color[rgb]{0,0,0}\makebox(0,0)[lb]{\smash{-5}}}%
    \put(0.30555004,0.08716212){\color[rgb]{0,0,0}\makebox(0,0)[lb]{\smash{0}}}%
    \put(0.36624979,0.11314256){\color[rgb]{0,0,0}\makebox(0,0)[lb]{\smash{5}}}%
    \put(0.41894317,0.13176886){\color[rgb]{0,0,0}\makebox(0,0)[lb]{\smash{10}}}%
    \put(0.31666157,0.07082816){\color[rgb]{0,0,0}\makebox(0,0)[lb]{\smash{$x$}}}%
    \put(0.4887021,0.19416281){\color[rgb]{0,0,0}\rotatebox{90}{\makebox(0,0)[lb]{\smash{$\chi(z,x)$}}}}%
    \put(0.77308147,0.03120211){\color[rgb]{0,0,0}\makebox(0,0)[lb]{\smash{-10}}}%
    \put(0.83933457,0.0614315){\color[rgb]{0,0,0}\makebox(0,0)[lb]{\smash{-5}}}%
    \put(0.89978848,0.08937335){\color[rgb]{0,0,0}\makebox(0,0)[lb]{\smash{0}}}%
    \put(0.94937193,0.11813287){\color[rgb]{0,0,0}\makebox(0,0)[lb]{\smash{5}}}%
    \put(0.97844313,0.14092779){\color[rgb]{0,0,0}\makebox(0,0)[lb]{\smash{10}}}%
    \put(0.91090005,0.07026031){\color[rgb]{0,0,0}\makebox(0,0)[lb]{\smash{$x$}}}%
    \put(0.98902853,0.1630713){\color[rgb]{0,0,0}\makebox(0,0)[lb]{\smash{0}}}%
    \put(0.99450638,0.20669492){\color[rgb]{0,0,0}\makebox(0,0)[lb]{\smash{2}}}%
    \put(1.0003858,0.25685501){\color[rgb]{0,0,0}\makebox(0,0)[lb]{\smash{4}}}%
    \put(1.04389265,0.1791429){\color[rgb]{0,0,0}\rotatebox{90}{\makebox(0,0)[lb]{\smash{$\phi(z,x)$}}}}%
    \put(0.72448844,0.02482705){\color[rgb]{0,0,0}\makebox(0,0)[lb]{\smash{1.0}}}%
    \put(0.61844461,0.09884553){\color[rgb]{0,0,0}\makebox(0,0)[lb]{\smash{0.5}}}%
    \put(0.5296376,0.15742319){\color[rgb]{0,0,0}\makebox(0,0)[lb]{\smash{0.0}}}%
    \put(0.59563487,0.07936466){\color[rgb]{0,0,0}\makebox(0,0)[lb]{\smash{$z$}}}%
  \end{picture}%
\endgroup%

%% file: chargedensity.pdf_tex
\begingroup%
  \makeatletter%
  \providecommand\color[2][]{%
    \errmessage{(Inkscape) Color is used for the text in Inkscape, but the package 'color.sty' is not loaded}%
    \renewcommand\color[2][]{}%
  }%
  \providecommand\transparent[1]{%
    \errmessage{(Inkscape) Transparency is used (non-zero) for the text in Inkscape, but the package 'transparent.sty' is not loaded}%
    \renewcommand\transparent[1]{}%
  }%
  \providecommand\rotatebox[2]{#2}%
  \ifx\svgwidth\undefined%
    \setlength{\unitlength}{422.30400391bp}%
    \ifx\svgscale\undefined%
      \relax%
    \else%
      \setlength{\unitlength}{\unitlength * \real{\svgscale}}%
    \fi%
  \else%
    \setlength{\unitlength}{\svgwidth}%
  \fi%
  \global\let\svgwidth\undefined%
  \global\let\svgscale\undefined%
  \makeatother%
  \begin{picture}(1,0.64146361)%
    \put(0,0){\includegraphics[width=\unitlength]{chargedensity.pdf}}%
    \put(0.01624368,-0.0073123){\color[rgb]{0,0,0}\makebox(0,0)[lb]{\smash{-}}}%
    \put(0.03518738,-0.0073123){\color[rgb]{0,0,0}\makebox(0,0)[lb]{\smash{10}}}%
    \put(0.25836101,-0.0073123){\color[rgb]{0,0,0}\makebox(0,0)[lb]{\smash{-}}}%
    \put(0.27730471,-0.0073123){\color[rgb]{0,0,0}\makebox(0,0)[lb]{\smash{5}}}%
    \put(0.50331988,-0.0073123){\color[rgb]{0,0,0}\makebox(0,0)[lb]{\smash{0}}}%
    \put(0.7388069,-0.0073123){\color[rgb]{0,0,0}\makebox(0,0)[lb]{\smash{5}}}%
    \put(0.96766365,-0.0073123){\color[rgb]{0,0,0}\makebox(0,0)[lb]{\smash{10}}}%
    \put(-0.01060847,0.0244184){\color[rgb]{0,0,0}\makebox(0,0)[lb]{\smash{0}}}%
    \put(-0.01074108,0.15896112){\color[rgb]{0,0,0}\makebox(0,0)[lb]{\smash{2}}}%
    \put(-0.01045692,0.29350383){\color[rgb]{0,0,0}\makebox(0,0)[lb]{\smash{4}}}%
    \put(-0.01094946,0.42804653){\color[rgb]{0,0,0}\makebox(0,0)[lb]{\smash{6}}}%
    \put(-0.01180192,0.56258924){\color[rgb]{0,0,0}\makebox(0,0)[lb]{\smash{8}}}%
    \put(0.49106042,-0.04533185){\color[rgb]{0,0,0}\makebox(0,0)[lb]{\smash{$x$}}}%
    \put(-0.04472798,0.27208506){\color[rgb]{0,0,0}\rotatebox{90}{\makebox(0,0)[lb]{\smash{$\bar\rho(x)/T^2$}}}}%
  \end{picture}%
\endgroup%

%% file: moshe.pdf_tex
\begingroup%
  \makeatletter%
  \providecommand\color[2][]{%
    \errmessage{(Inkscape) Color is used for the text in Inkscape, but the package 'color.sty' is not loaded}%
    \renewcommand\color[2][]{}%
  }%
  \providecommand\transparent[1]{%
    \errmessage{(Inkscape) Transparency is used (non-zero) for the text in Inkscape, but the package 'transparent.sty' is not loaded}%
    \renewcommand\transparent[1]{}%
  }%
  \providecommand\rotatebox[2]{#2}%
  \ifx\svgwidth\undefined%
    \setlength{\unitlength}{499.74609375bp}%
    \ifx\svgscale\undefined%
      \relax%
    \else%
      \setlength{\unitlength}{\unitlength * \real{\svgscale}}%
    \fi%
  \else%
    \setlength{\unitlength}{\svgwidth}%
  \fi%
  \global\let\svgwidth\undefined%
  \global\let\svgscale\undefined%
  \makeatother%
  \begin{picture}(1,0.62813465)%
    \put(0,0){\includegraphics[width=\unitlength]{moshe.pdf}}%
    \put(0.07245637,-0.00938074){\color[rgb]{0,0,0}\makebox(0,0)[lb]{\smash{1.5}}}%
    \put(0.22529216,-0.00938074){\color[rgb]{0,0,0}\makebox(0,0)[lb]{\smash{2.0}}}%
    \put(0.37812793,-0.00938074){\color[rgb]{0,0,0}\makebox(0,0)[lb]{\smash{2.5}}}%
    \put(0.53096372,-0.00938074){\color[rgb]{0,0,0}\makebox(0,0)[lb]{\smash{3.0}}}%
    \put(0.68379951,-0.00938074){\color[rgb]{0,0,0}\makebox(0,0)[lb]{\smash{3.5}}}%
    \put(0.8366353,-0.00938074){\color[rgb]{0,0,0}\makebox(0,0)[lb]{\smash{4.0}}}%
    \put(-0.00459296,0.08068461){\color[rgb]{0,0,0}\makebox(0,0)[lb]{\smash{-1}}}%
    \put(0.00352598,0.16144055){\color[rgb]{0,0,0}\makebox(0,0)[lb]{\smash{0}}}%
    \put(0.0008206,0.24219649){\color[rgb]{0,0,0}\makebox(0,0)[lb]{\smash{1}}}%
    \put(0.00341392,0.32295243){\color[rgb]{0,0,0}\makebox(0,0)[lb]{\smash{2}}}%
    \put(0.00251747,0.40370836){\color[rgb]{0,0,0}\makebox(0,0)[lb]{\smash{3}}}%
    \put(0.00365404,0.48446431){\color[rgb]{0,0,0}\makebox(0,0)[lb]{\smash{4}}}%
    \put(0.0027896,0.56522024){\color[rgb]{0,0,0}\makebox(0,0)[lb]{\smash{5}}}%
    \put(-0.044417,0.27909659){\color[rgb]{0,0,0}\rotatebox{90}{\makebox(0,0)[lb]{\smash{$\log \frac{\bar\rho(x)}{T^2}$}}}}%
    \put(0.4568017,-0.03606328){\color[rgb]{0,0,0}\makebox(0,0)[lb]{\smash{$\log \mu$}}}%
  \end{picture}%
\endgroup%

%% file: sigmawxXandY.pdf_tex
\begingroup%
  \makeatletter%
  \providecommand\color[2][]{%
    \errmessage{(Inkscape) Color is used for the text in Inkscape, but the package 'color.sty' is not loaded}%
    \renewcommand\color[2][]{}%
  }%
  \providecommand\transparent[1]{%
    \errmessage{(Inkscape) Transparency is used (non-zero) for the text in Inkscape, but the package 'transparent.sty' is not loaded}%
    \renewcommand\transparent[1]{}%
  }%
  \providecommand\rotatebox[2]{#2}%
  \ifx\svgwidth\undefined%
    \setlength{\unitlength}{1141.91748047bp}%
    \ifx\svgscale\undefined%
      \relax%
    \else%
      \setlength{\unitlength}{\unitlength * \real{\svgscale}}%
    \fi%
  \else%
    \setlength{\unitlength}{\svgwidth}%
  \fi%
  \global\let\svgwidth\undefined%
  \global\let\svgscale\undefined%
  \makeatother%
  \begin{picture}(1,0.27819074)%
    \put(0,0){\includegraphics[width=\unitlength]{sigmawxXandY.pdf}}%
    \put(0.32535105,0.26333676){\color[rgb]{0,0,0}\makebox(0,0)[lb]{\smash{$x$}}}%
    \put(0.46635602,0.21523295){\color[rgb]{0,0,0}\rotatebox{-90}{\makebox(0,0)[lb]{\smash{Re $\sigma_x(x,\omega)$}}}}%
    \put(0.31731659,0.04243364){\color[rgb]{0,0,0}\makebox(0,0)[lb]{\smash{$\omega$}}}%
    \put(0.88357713,0.26119069){\color[rgb]{0,0,0}\makebox(0,0)[lb]{\smash{$x$}}}%
    \put(0.87797252,0.03267877){\color[rgb]{0,0,0}\makebox(0,0)[lb]{\smash{$\omega$}}}%
    \put(1.04828595,0.21545197){\color[rgb]{0,0,0}\rotatebox{-90}{\makebox(0,0)[lb]{\smash{Re $\sigma_y(x,\omega)$}}}}%
  \end{picture}%
\endgroup%

%% file: sigmaACyvshomogALL.pdf_tex
\begingroup%
  \makeatletter%
  \providecommand\color[2][]{%
    \errmessage{(Inkscape) Color is used for the text in Inkscape, but the package 'color.sty' is not loaded}%
    \renewcommand\color[2][]{}%
  }%
  \providecommand\transparent[1]{%
    \errmessage{(Inkscape) Transparency is used (non-zero) for the text in Inkscape, but the package 'transparent.sty' is not loaded}%
    \renewcommand\transparent[1]{}%
  }%
  \providecommand\rotatebox[2]{#2}%
  \ifx\svgwidth\undefined%
    \setlength{\unitlength}{1961.903125bp}%
    \ifx\svgscale\undefined%
      \relax%
    \else%
      \setlength{\unitlength}{\unitlength * \real{\svgscale}}%
    \fi%
  \else%
    \setlength{\unitlength}{\svgwidth}%
  \fi%
  \global\let\svgwidth\undefined%
  \global\let\svgscale\undefined%
  \makeatother%
  \begin{picture}(1,0.19822359)%
    \put(0,0){\includegraphics[width=\unitlength]{sigmaACyvshomogALL.pdf}}%
    \put(0.72472241,0.00379761){\color[rgb]{0,0,0}\makebox(0,0)[lb]{\smash{0}}}%
    \put(0.79144658,0.00379761){\color[rgb]{0,0,0}\makebox(0,0)[lb]{\smash{4}}}%
    \put(0.85817076,0.00379761){\color[rgb]{0,0,0}\makebox(0,0)[lb]{\smash{8}}}%
    \put(0.92346775,0.00379761){\color[rgb]{0,0,0}\makebox(0,0)[lb]{\smash{12}}}%
    \put(0.70594296,0.04863378){\color[rgb]{0,0,0}\makebox(0,0)[lb]{\smash{1}}}%
    \put(0.70594296,0.07739087){\color[rgb]{0,0,0}\makebox(0,0)[lb]{\smash{2}}}%
    \put(0.70594296,0.10614797){\color[rgb]{0,0,0}\makebox(0,0)[lb]{\smash{3}}}%
    \put(0.70594296,0.13490507){\color[rgb]{0,0,0}\makebox(0,0)[lb]{\smash{4}}}%
    \put(0.70594296,0.16366216){\color[rgb]{0,0,0}\makebox(0,0)[lb]{\smash{5}}}%
    \put(0.39966656,0.00421335){\color[rgb]{0,0,0}\makebox(0,0)[lb]{\smash{0}}}%
    \put(0.46722855,0.00421335){\color[rgb]{0,0,0}\makebox(0,0)[lb]{\smash{4}}}%
    \put(0.53479054,0.00421335){\color[rgb]{0,0,0}\makebox(0,0)[lb]{\smash{8}}}%
    \put(0.60092534,0.00421335){\color[rgb]{0,0,0}\makebox(0,0)[lb]{\smash{12}}}%
    \put(0.38475243,0.03733122){\color[rgb]{0,0,0}\makebox(0,0)[rb]{\smash{0.5}}}%
    \put(0.38455263,0.06446677){\color[rgb]{0,0,0}\makebox(0,0)[rb]{\smash{1.0}}}%
    \put(0.38439768,0.09160232){\color[rgb]{0,0,0}\makebox(0,0)[rb]{\smash{1.5}}}%
    \put(0.38488292,0.11873787){\color[rgb]{0,0,0}\makebox(0,0)[rb]{\smash{2.0}}}%
    \put(0.38472797,0.14587342){\color[rgb]{0,0,0}\makebox(0,0)[rb]{\smash{2.5}}}%
    \put(0.38482991,0.17300897){\color[rgb]{0,0,0}\makebox(0,0)[rb]{\smash{3.0}}}%
    \put(0.00382521,0.0686556){\color[rgb]{0,0,0}\rotatebox{90}{\makebox(0,0)[lb]{\smash{Re $\sigma(\omega)$}}}}%
    \put(0.05868823,0.00476611){\color[rgb]{0,0,0}\makebox(0,0)[lb]{\smash{0}}}%
    \put(0.12589108,0.00476611){\color[rgb]{0,0,0}\makebox(0,0)[lb]{\smash{4}}}%
    \put(0.19309393,0.00476611){\color[rgb]{0,0,0}\makebox(0,0)[lb]{\smash{8}}}%
    \put(0.25859807,0.00476611){\color[rgb]{0,0,0}\makebox(0,0)[lb]{\smash{12}}}%
    \put(0.01713716,0.03971672){\color[rgb]{0,0,0}\makebox(0,0)[lb]{\smash{0.6}}}%
    \put(0.04340096,0.07072508){\color[rgb]{0,0,0}\makebox(0,0)[rb]{\smash{0.8}}}%
    \put(0.04312917,0.10173344){\color[rgb]{0,0,0}\makebox(0,0)[rb]{\smash{1.0}}}%
    \put(0.04312431,0.1327418){\color[rgb]{0,0,0}\makebox(0,0)[rb]{\smash{1.2}}}%
    \put(0.04310975,0.16375016){\color[rgb]{0,0,0}\makebox(0,0)[rb]{\smash{1.4}}}%
    \put(0.3452964,0.06838433){\color[rgb]{0,0,0}\rotatebox{90}{\makebox(0,0)[lb]{\smash{Re $\sigma(\omega)$}}}}%
    \put(0.69026757,0.06838433){\color[rgb]{0,0,0}\rotatebox{90}{\makebox(0,0)[lb]{\smash{Re $\sigma(\omega)$}}}}%
    \put(0.12147318,0.19199268){\color[rgb]{0,0,0}\makebox(0,0)[lb]{\smash{$M=3$, $\mu=2$}}}%
    \put(0.78384845,0.19439838){\color[rgb]{0,0,0}\makebox(0,0)[lb]{\smash{$M=7$, $\mu=6$}}}%
    \put(0.45763459,0.19439838){\color[rgb]{0,0,0}\makebox(0,0)[lb]{\smash{$M=5$, $\mu=4$}}}%
    \put(0.29818193,0.00226978){\color[rgb]{0,0,0}\makebox(0,0)[lb]{\smash{$\omega$}}}%
    \put(0.64944992,0.00171629){\color[rgb]{0,0,0}\makebox(0,0)[lb]{\smash{$\omega$}}}%
    \put(0.96406954,0.00253183){\color[rgb]{0,0,0}\makebox(0,0)[lb]{\smash{$\omega$}}}%
  \end{picture}%
\endgroup%

%% file: sigmaACyvshomogGRADIENTS.pdf_tex
\begingroup%
  \makeatletter%
  \providecommand\color[2][]{%
    \errmessage{(Inkscape) Color is used for the text in Inkscape, but the package 'color.sty' is not loaded}%
    \renewcommand\color[2][]{}%
  }%
  \providecommand\transparent[1]{%
    \errmessage{(Inkscape) Transparency is used (non-zero) for the text in Inkscape, but the package 'transparent.sty' is not loaded}%
    \renewcommand\transparent[1]{}%
  }%
  \providecommand\rotatebox[2]{#2}%
  \ifx\svgwidth\undefined%
    \setlength{\unitlength}{1189.75546875bp}%
    \ifx\svgscale\undefined%
      \relax%
    \else%
      \setlength{\unitlength}{\unitlength * \real{\svgscale}}%
    \fi%
  \else%
    \setlength{\unitlength}{\svgwidth}%
  \fi%
  \global\let\svgwidth\undefined%
  \global\let\svgscale\undefined%
  \makeatother%
  \begin{picture}(1,0.30854095)%
    \put(0,0){\includegraphics[width=\unitlength]{sigmaACyvshomogGRADIENTS.pdf}}%
    \put(0.57215457,0.02620317){\color[rgb]{0,0,0}\makebox(0,0)[lb]{\smash{0}}}%
    \put(0.63767906,0.02620317){\color[rgb]{0,0,0}\makebox(0,0)[lb]{\smash{2}}}%
    \put(0.70320356,0.02620317){\color[rgb]{0,0,0}\makebox(0,0)[lb]{\smash{4}}}%
    \put(0.76872806,0.02620317){\color[rgb]{0,0,0}\makebox(0,0)[lb]{\smash{6}}}%
    \put(0.83425256,0.02620317){\color[rgb]{0,0,0}\makebox(0,0)[lb]{\smash{8}}}%
    \put(0.89790328,0.02620317){\color[rgb]{0,0,0}\makebox(0,0)[lb]{\smash{10}}}%
    \put(0.96342777,0.02620317){\color[rgb]{0,0,0}\makebox(0,0)[lb]{\smash{12}}}%
    \put(0.53185496,0.0689392){\color[rgb]{0,0,0}\makebox(0,0)[lb]{\smash{0.6}}}%
    \put(0.53185496,0.12088427){\color[rgb]{0,0,0}\makebox(0,0)[lb]{\smash{0.8}}}%
    \put(0.53185496,0.17282934){\color[rgb]{0,0,0}\makebox(0,0)[lb]{\smash{1.0}}}%
    \put(0.53185496,0.22477442){\color[rgb]{0,0,0}\makebox(0,0)[lb]{\smash{1.2}}}%
    \put(0.53185496,0.27671949){\color[rgb]{0,0,0}\makebox(0,0)[lb]{\smash{1.4}}}%
    \put(0.79046259,0.00169809){\color[rgb]{0,0,0}\makebox(0,0)[rb]{\smash{$\omega$}}}%
    \put(0.51958445,0.22519586){\color[rgb]{0,0,0}\rotatebox{90}{\makebox(0,0)[rb]{\smash{Re $\sigma(\omega)$}}}}%
    \put(0.07168309,0.02621614){\color[rgb]{0,0,0}\makebox(0,0)[lb]{\smash{-10}}}%
    \put(0.17566382,0.02621614){\color[rgb]{0,0,0}\makebox(0,0)[lb]{\smash{-5}}}%
    \put(0.27461006,0.02621614){\color[rgb]{0,0,0}\makebox(0,0)[lb]{\smash{0}}}%
    \put(0.37643315,0.02621614){\color[rgb]{0,0,0}\makebox(0,0)[lb]{\smash{5}}}%
    \put(0.47609861,0.02621614){\color[rgb]{0,0,0}\makebox(0,0)[lb]{\smash{10}}}%
    \put(0.02479417,0.11264138){\color[rgb]{0,0,0}\makebox(0,0)[lb]{\smash{2.0}}}%
    \put(0.02479417,0.18310636){\color[rgb]{0,0,0}\makebox(0,0)[lb]{\smash{2.5}}}%
    \put(0.02479417,0.25357134){\color[rgb]{0,0,0}\makebox(0,0)[lb]{\smash{3.0}}}%
    \put(0.28461949,0.00338747){\color[rgb]{0,0,0}\makebox(0,0)[rb]{\smash{$x$}}}%
    \put(0.0103422,0.21440057){\color[rgb]{0,0,0}\rotatebox{90}{\makebox(0,0)[rb]{\smash{$\bar\rho(x)/T^2$}}}}%
  \end{picture}%
\endgroup%

%% file: sigmasinterface.pdf_tex
\begingroup%
  \makeatletter%
  \providecommand\color[2][]{%
    \errmessage{(Inkscape) Color is used for the text in Inkscape, but the package 'color.sty' is not loaded}%
    \renewcommand\color[2][]{}%
  }%
  \providecommand\transparent[1]{%
    \errmessage{(Inkscape) Transparency is used (non-zero) for the text in Inkscape, but the package 'transparent.sty' is not loaded}%
    \renewcommand\transparent[1]{}%
  }%
  \providecommand\rotatebox[2]{#2}%
  \ifx\svgwidth\undefined%
    \setlength{\unitlength}{1128.39511719bp}%
    \ifx\svgscale\undefined%
      \relax%
    \else%
      \setlength{\unitlength}{\unitlength * \real{\svgscale}}%
    \fi%
  \else%
    \setlength{\unitlength}{\svgwidth}%
  \fi%
  \global\let\svgwidth\undefined%
  \global\let\svgscale\undefined%
  \makeatother%
  \begin{picture}(1,1.09904216)%
    \put(0,0){\includegraphics[width=\unitlength]{sigmasinterface.pdf}}%
    \put(0.59444637,0.01906648){\color[rgb]{0,0,0}\makebox(0,0)[lb]{\smash{0}}}%
    \put(0.6472684,0.01906648){\color[rgb]{0,0,0}\makebox(0,0)[lb]{\smash{2}}}%
    \put(0.70009044,0.01906648){\color[rgb]{0,0,0}\makebox(0,0)[lb]{\smash{4}}}%
    \put(0.75291247,0.01906648){\color[rgb]{0,0,0}\makebox(0,0)[lb]{\smash{6}}}%
    \put(0.80573451,0.01906648){\color[rgb]{0,0,0}\makebox(0,0)[lb]{\smash{8}}}%
    \put(0.85616554,0.01906648){\color[rgb]{0,0,0}\makebox(0,0)[lb]{\smash{10}}}%
    \put(0.90898757,0.01906648){\color[rgb]{0,0,0}\makebox(0,0)[lb]{\smash{12}}}%
    \put(0.9618096,0.01906648){\color[rgb]{0,0,0}\makebox(0,0)[lb]{\smash{14}}}%
    \put(0.58388433,0.09965777){\color[rgb]{0,0,0}\makebox(0,0)[rb]{\smash{-3}}}%
    \put(0.5842669,0.16348509){\color[rgb]{0,0,0}\makebox(0,0)[rb]{\smash{-2}}}%
    \put(0.58316021,0.22731239){\color[rgb]{0,0,0}\makebox(0,0)[rb]{\smash{-1}}}%
    \put(0.58431472,0.29113968){\color[rgb]{0,0,0}\makebox(0,0)[rb]{\smash{0}}}%
    \put(0.24307485,0.00179043){\color[rgb]{0,0,0}\makebox(0,0)[lb]{\smash{$\omega$}}}%
    \put(0.77892791,0.00301607){\color[rgb]{0,0,0}\makebox(0,0)[lb]{\smash{$\omega$}}}%
    \put(0.00665085,0.14113319){\color[rgb]{0,0,0}\rotatebox{90}{\makebox(0,0)[lb]{\smash{Re $\sigma(\omega)$}}}}%
    \put(0.54208346,0.14190248){\color[rgb]{0,0,0}\rotatebox{90}{\makebox(0,0)[lb]{\smash{Im $\sigma(\omega)$}}}}%
    \put(0.05615422,0.01906492){\color[rgb]{0,0,0}\makebox(0,0)[lb]{\smash{0}}}%
    \put(0.10983652,0.01906492){\color[rgb]{0,0,0}\makebox(0,0)[lb]{\smash{2}}}%
    \put(0.16351882,0.01906492){\color[rgb]{0,0,0}\makebox(0,0)[lb]{\smash{4}}}%
    \put(0.21720112,0.01906492){\color[rgb]{0,0,0}\makebox(0,0)[lb]{\smash{6}}}%
    \put(0.27088342,0.01906492){\color[rgb]{0,0,0}\makebox(0,0)[lb]{\smash{8}}}%
    \put(0.32217548,0.01906492){\color[rgb]{0,0,0}\makebox(0,0)[lb]{\smash{10}}}%
    \put(0.37585778,0.01906492){\color[rgb]{0,0,0}\makebox(0,0)[lb]{\smash{12}}}%
    \put(0.42954008,0.01906492){\color[rgb]{0,0,0}\makebox(0,0)[lb]{\smash{14}}}%
    \put(0.04584072,0.10306935){\color[rgb]{0,0,0}\makebox(0,0)[rb]{\smash{2}}}%
    \put(0.04594316,0.16832014){\color[rgb]{0,0,0}\makebox(0,0)[rb]{\smash{4}}}%
    \put(0.0457656,0.23357093){\color[rgb]{0,0,0}\makebox(0,0)[rb]{\smash{6}}}%
    \put(0.04545828,0.29882172){\color[rgb]{0,0,0}\makebox(0,0)[rb]{\smash{8}}}%
    \put(0.24307472,0.38748123){\color[rgb]{0,0,0}\makebox(0,0)[lb]{\smash{$\omega$}}}%
    \put(0.77892813,0.38748123){\color[rgb]{0,0,0}\makebox(0,0)[lb]{\smash{$\omega$}}}%
    \put(0.00665085,0.52747455){\color[rgb]{0,0,0}\rotatebox{90}{\makebox(0,0)[lb]{\smash{Re $\sigma(\omega)$}}}}%
    \put(0.54098373,0.52186016){\color[rgb]{0,0,0}\rotatebox{90}{\makebox(0,0)[lb]{\smash{Im $\sigma(\omega)$}}}}%
    \put(0.59819532,0.40189585){\color[rgb]{0,0,0}\makebox(0,0)[lb]{\smash{0}}}%
    \put(0.65057998,0.40189585){\color[rgb]{0,0,0}\makebox(0,0)[lb]{\smash{2}}}%
    \put(0.70296465,0.40189585){\color[rgb]{0,0,0}\makebox(0,0)[lb]{\smash{4}}}%
    \put(0.75534931,0.40189585){\color[rgb]{0,0,0}\makebox(0,0)[lb]{\smash{6}}}%
    \put(0.80773397,0.40189585){\color[rgb]{0,0,0}\makebox(0,0)[lb]{\smash{8}}}%
    \put(0.85811932,0.40189585){\color[rgb]{0,0,0}\makebox(0,0)[lb]{\smash{10}}}%
    \put(0.91050398,0.40189585){\color[rgb]{0,0,0}\makebox(0,0)[lb]{\smash{12}}}%
    \put(0.96288864,0.40189585){\color[rgb]{0,0,0}\makebox(0,0)[lb]{\smash{14}}}%
    \put(0.58697527,0.43754095){\color[rgb]{0,0,0}\makebox(0,0)[rb]{\smash{-2.0}}}%
    \put(0.58629551,0.49416893){\color[rgb]{0,0,0}\makebox(0,0)[rb]{\smash{-1.5}}}%
    \put(0.58651258,0.55079691){\color[rgb]{0,0,0}\makebox(0,0)[rb]{\smash{-1.0}}}%
    \put(0.58679248,0.60742489){\color[rgb]{0,0,0}\makebox(0,0)[rb]{\smash{-0.5}}}%
    \put(0.58700955,0.66405287){\color[rgb]{0,0,0}\makebox(0,0)[rb]{\smash{0.0}}}%
    \put(0.05577756,0.40190179){\color[rgb]{0,0,0}\makebox(0,0)[lb]{\smash{0}}}%
    \put(0.1095352,0.40190179){\color[rgb]{0,0,0}\makebox(0,0)[lb]{\smash{2}}}%
    \put(0.16329285,0.40190179){\color[rgb]{0,0,0}\makebox(0,0)[lb]{\smash{4}}}%
    \put(0.2170505,0.40190179){\color[rgb]{0,0,0}\makebox(0,0)[lb]{\smash{6}}}%
    \put(0.27080814,0.40190179){\color[rgb]{0,0,0}\makebox(0,0)[lb]{\smash{8}}}%
    \put(0.32236623,0.40190179){\color[rgb]{0,0,0}\makebox(0,0)[lb]{\smash{10}}}%
    \put(0.37612386,0.40190179){\color[rgb]{0,0,0}\makebox(0,0)[lb]{\smash{12}}}%
    \put(0.42988151,0.40190179){\color[rgb]{0,0,0}\makebox(0,0)[lb]{\smash{14}}}%
    \put(0.03466113,0.46618301){\color[rgb]{0,0,0}\makebox(0,0)[lb]{\smash{1}}}%
    \put(0.04560482,0.53053372){\color[rgb]{0,0,0}\makebox(0,0)[rb]{\smash{2}}}%
    \put(0.04525289,0.59488444){\color[rgb]{0,0,0}\makebox(0,0)[rb]{\smash{3}}}%
    \put(0.04569909,0.65923515){\color[rgb]{0,0,0}\makebox(0,0)[rb]{\smash{4}}}%
    \put(0.24307471,0.77032576){\color[rgb]{0,0,0}\makebox(0,0)[lb]{\smash{$\omega$}}}%
    \put(0.77892813,0.77229506){\color[rgb]{0,0,0}\makebox(0,0)[lb]{\smash{$\omega$}}}%
    \put(0.00665076,0.90122833){\color[rgb]{0,0,0}\rotatebox{90}{\makebox(0,0)[lb]{\smash{Re $\sigma(\omega)$}}}}%
    \put(0.54098372,0.90470478){\color[rgb]{0,0,0}\rotatebox{90}{\makebox(0,0)[lb]{\smash{Im $\sigma(\omega)$}}}}%
    \put(0.5996815,0.78474899){\color[rgb]{0,0,0}\makebox(0,0)[lb]{\smash{0}}}%
    \put(0.65184417,0.78474899){\color[rgb]{0,0,0}\makebox(0,0)[lb]{\smash{2}}}%
    \put(0.70400684,0.78474899){\color[rgb]{0,0,0}\makebox(0,0)[lb]{\smash{4}}}%
    \put(0.75616952,0.78474899){\color[rgb]{0,0,0}\makebox(0,0)[lb]{\smash{6}}}%
    \put(0.80833219,0.78474899){\color[rgb]{0,0,0}\makebox(0,0)[lb]{\smash{8}}}%
    \put(0.8582979,0.78474899){\color[rgb]{0,0,0}\makebox(0,0)[lb]{\smash{10}}}%
    \put(0.91046057,0.78474899){\color[rgb]{0,0,0}\makebox(0,0)[lb]{\smash{12}}}%
    \put(0.96262325,0.78474899){\color[rgb]{0,0,0}\makebox(0,0)[lb]{\smash{14}}}%
    \put(0.55233125,0.83791396){\color[rgb]{0,0,0}\makebox(0,0)[lb]{\smash{-0.6}}}%
    \put(0.55233125,0.90052297){\color[rgb]{0,0,0}\makebox(0,0)[lb]{\smash{-0.4}}}%
    \put(0.55233125,0.96313198){\color[rgb]{0,0,0}\makebox(0,0)[lb]{\smash{-0.2}}}%
    \put(0.55233125,1.02574099){\color[rgb]{0,0,0}\makebox(0,0)[lb]{\smash{0.0}}}%
    \put(0.06179719,0.78474775){\color[rgb]{0,0,0}\makebox(0,0)[lb]{\smash{0}}}%
    \put(0.11476609,0.78474775){\color[rgb]{0,0,0}\makebox(0,0)[lb]{\smash{2}}}%
    \put(0.167735,0.78474775){\color[rgb]{0,0,0}\makebox(0,0)[lb]{\smash{4}}}%
    \put(0.22070391,0.78474775){\color[rgb]{0,0,0}\makebox(0,0)[lb]{\smash{6}}}%
    \put(0.27367282,0.78474775){\color[rgb]{0,0,0}\makebox(0,0)[lb]{\smash{8}}}%
    \put(0.32444178,0.78474775){\color[rgb]{0,0,0}\makebox(0,0)[lb]{\smash{10}}}%
    \put(0.37741069,0.78474775){\color[rgb]{0,0,0}\makebox(0,0)[lb]{\smash{12}}}%
    \put(0.43037959,0.78474775){\color[rgb]{0,0,0}\makebox(0,0)[lb]{\smash{14}}}%
    \put(0.05009321,0.83696257){\color[rgb]{0,0,0}\makebox(0,0)[rb]{\smash{0.8}}}%
    \put(0.04974121,0.87731332){\color[rgb]{0,0,0}\makebox(0,0)[rb]{\smash{1.0}}}%
    \put(0.04973493,0.91766406){\color[rgb]{0,0,0}\makebox(0,0)[rb]{\smash{1.2}}}%
    \put(0.04971607,0.95801481){\color[rgb]{0,0,0}\makebox(0,0)[rb]{\smash{1.4}}}%
    \put(0.04969093,0.99836555){\color[rgb]{0,0,0}\makebox(0,0)[rb]{\smash{1.6}}}%
    \put(0.04954636,1.0387163){\color[rgb]{0,0,0}\makebox(0,0)[rb]{\smash{1.8}}}%
    \put(0.19192046,1.0923914){\color[rgb]{0,0,0}\makebox(0,0)[lb]{\smash{$M=3$, $\mu=2$}}}%
    \put(0.71531327,1.09214405){\color[rgb]{0,0,0}\makebox(0,0)[lb]{\smash{$M=3$, $\mu=2$}}}%
    \put(0.19309432,0.70991836){\color[rgb]{0,0,0}\makebox(0,0)[lb]{\smash{$M=5$, $\mu=4$}}}%
    \put(0.19309432,0.32707378){\color[rgb]{0,0,0}\makebox(0,0)[lb]{\smash{$M=7$, $\mu=6$}}}%
    \put(0.71729445,0.71034987){\color[rgb]{0,0,0}\makebox(0,0)[lb]{\smash{$M=5$, $\mu=4$}}}%
    \put(0.73258729,0.32662029){\color[rgb]{0,0,0}\makebox(0,0)[lb]{\smash{$M=7$, $\mu=6$}}}%
  \end{picture}%
\endgroup%

%% file: comparesigmadcS.pdf_tex
\begingroup%
  \makeatletter%
  \providecommand\color[2][]{%
    \errmessage{(Inkscape) Color is used for the text in Inkscape, but the package 'color.sty' is not loaded}%
    \renewcommand\color[2][]{}%
  }%
  \providecommand\transparent[1]{%
    \errmessage{(Inkscape) Transparency is used (non-zero) for the text in Inkscape, but the package 'transparent.sty' is not loaded}%
    \renewcommand\transparent[1]{}%
  }%
  \providecommand\rotatebox[2]{#2}%
  \ifx\svgwidth\undefined%
    \setlength{\unitlength}{835.37734375bp}%
    \ifx\svgscale\undefined%
      \relax%
    \else%
      \setlength{\unitlength}{\unitlength * \real{\svgscale}}%
    \fi%
  \else%
    \setlength{\unitlength}{\svgwidth}%
  \fi%
  \global\let\svgwidth\undefined%
  \global\let\svgscale\undefined%
  \makeatother%
  \begin{picture}(1,0.31988073)%
    \put(0,0){\includegraphics[width=\unitlength]{comparesigmadcS.pdf}}%
    \put(0.57646267,0.01927925){\color[rgb]{0,0,0}\makebox(0,0)[lb]{\smash{-}}}%
    \put(0.58406483,0.01927925){\color[rgb]{0,0,0}\makebox(0,0)[lb]{\smash{10}}}%
    \put(0.68058948,0.01927925){\color[rgb]{0,0,0}\makebox(0,0)[lb]{\smash{-}}}%
    \put(0.68819164,0.01927925){\color[rgb]{0,0,0}\makebox(0,0)[lb]{\smash{5}}}%
    \put(0.78585661,0.01927925){\color[rgb]{0,0,0}\makebox(0,0)[lb]{\smash{0}}}%
    \put(0.88732266,0.01927925){\color[rgb]{0,0,0}\makebox(0,0)[lb]{\smash{5}}}%
    \put(0.98612796,0.01927925){\color[rgb]{0,0,0}\makebox(0,0)[lb]{\smash{10}}}%
    \put(0.56084858,0.05757544){\color[rgb]{0,0,0}\makebox(0,0)[rb]{\smash{2}}}%
    \put(0.56042286,0.09842019){\color[rgb]{0,0,0}\makebox(0,0)[rb]{\smash{3}}}%
    \put(0.56096262,0.13926494){\color[rgb]{0,0,0}\makebox(0,0)[rb]{\smash{4}}}%
    \put(0.5605521,0.18010969){\color[rgb]{0,0,0}\makebox(0,0)[rb]{\smash{5}}}%
    \put(0.54735785,0.22095443){\color[rgb]{0,0,0}\makebox(0,0)[lb]{\smash{6}}}%
    \put(0.56072695,0.26179918){\color[rgb]{0,0,0}\makebox(0,0)[rb]{\smash{7}}}%
    \put(0.56042286,0.30264392){\color[rgb]{0,0,0}\makebox(0,0)[rb]{\smash{8}}}%
    \put(0.24867204,0.00118398){\color[rgb]{0,0,0}\makebox(0,0)[lb]{\smash{$x$}}}%
    \put(0.78441865,0.00075278){\color[rgb]{0,0,0}\makebox(0,0)[lb]{\smash{$x$}}}%
    \put(0.00898359,0.10320239){\color[rgb]{0,0,0}\rotatebox{90}{\makebox(0,0)[lb]{\smash{Re $\sigma(\omega=0,x)$}}}}%
    \put(0.53424419,0.10349371){\color[rgb]{0,0,0}\rotatebox{90}{\makebox(0,0)[lb]{\smash{Re $\sigma(\omega=0,x)$}}}}%
    \put(0.04384572,0.01928022){\color[rgb]{0,0,0}\makebox(0,0)[lb]{\smash{-}}}%
    \put(0.05151538,0.01928022){\color[rgb]{0,0,0}\makebox(0,0)[lb]{\smash{10}}}%
    \put(0.14797673,0.01928022){\color[rgb]{0,0,0}\makebox(0,0)[lb]{\smash{-}}}%
    \put(0.15564639,0.01928022){\color[rgb]{0,0,0}\makebox(0,0)[lb]{\smash{5}}}%
    \put(0.25325818,0.01928022){\color[rgb]{0,0,0}\makebox(0,0)[lb]{\smash{0}}}%
    \put(0.3547048,0.01928022){\color[rgb]{0,0,0}\makebox(0,0)[lb]{\smash{5}}}%
    \put(0.45346704,0.01928022){\color[rgb]{0,0,0}\makebox(0,0)[lb]{\smash{10}}}%
    \put(0.03401764,0.05765155){\color[rgb]{0,0,0}\makebox(0,0)[rb]{\smash{2}}}%
    \put(0.03358814,0.09848032){\color[rgb]{0,0,0}\makebox(0,0)[rb]{\smash{3}}}%
    \put(0.03413268,0.13930909){\color[rgb]{0,0,0}\makebox(0,0)[rb]{\smash{4}}}%
    \put(0.03371852,0.18013787){\color[rgb]{0,0,0}\makebox(0,0)[rb]{\smash{5}}}%
    \put(0.03393327,0.22096665){\color[rgb]{0,0,0}\makebox(0,0)[rb]{\smash{6}}}%
    \put(0.03389492,0.26179542){\color[rgb]{0,0,0}\makebox(0,0)[rb]{\smash{7}}}%
    \put(0.03358814,0.3026242){\color[rgb]{0,0,0}\makebox(0,0)[rb]{\smash{8}}}%
  \end{picture}%
\endgroup%

%% file: object.pdf_tex
\begingroup%
  \makeatletter%
  \providecommand\color[2][]{%
    \errmessage{(Inkscape) Color is used for the text in Inkscape, but the package 'color.sty' is not loaded}%
    \renewcommand\color[2][]{}%
  }%
  \providecommand\transparent[1]{%
    \errmessage{(Inkscape) Transparency is used (non-zero) for the text in Inkscape, but the package 'transparent.sty' is not loaded}%
    \renewcommand\transparent[1]{}%
  }%
  \providecommand\rotatebox[2]{#2}%
  \ifx\svgwidth\undefined%
    \setlength{\unitlength}{424.19760742bp}%
    \ifx\svgscale\undefined%
      \relax%
    \else%
      \setlength{\unitlength}{\unitlength * \real{\svgscale}}%
    \fi%
  \else%
    \setlength{\unitlength}{\svgwidth}%
  \fi%
  \global\let\svgwidth\undefined%
  \global\let\svgscale\undefined%
  \makeatother%
  \begin{picture}(1,0.44042679)%
    \put(0,0){\includegraphics[width=\unitlength]{object.pdf}}%
    \put(0.48320526,0.36221912){\color[rgb]{0,0,0}\makebox(0,0)[lb]{\smash{$\color{red}\sigma^{DC}_y$ }}}%
    \put(0.58881642,0.12459404){\color[rgb]{0,0,0}\makebox(0,0)[lb]{\smash{$\color{blue}\sigma^{DC}_x$ }}}%
    \put(0.96456178,0.12632381){\color[rgb]{0,0,0}\makebox(0,0)[lb]{\smash{$y$}}}%
    \put(0.92587086,0.05609402){\color[rgb]{0,0,0}\makebox(0,0)[lb]{\smash{$x$}}}%
  \end{picture}%
\endgroup%

%% file: tempsnew.pdf_tex
\begingroup%
  \makeatletter%
  \providecommand\color[2][]{%
    \errmessage{(Inkscape) Color is used for the text in Inkscape, but the package 'color.sty' is not loaded}%
    \renewcommand\color[2][]{}%
  }%
  \providecommand\transparent[1]{%
    \errmessage{(Inkscape) Transparency is used (non-zero) for the text in Inkscape, but the package 'transparent.sty' is not loaded}%
    \renewcommand\transparent[1]{}%
  }%
  \providecommand\rotatebox[2]{#2}%
  \ifx\svgwidth\undefined%
    \setlength{\unitlength}{1234.19375bp}%
    \ifx\svgscale\undefined%
      \relax%
    \else%
      \setlength{\unitlength}{\unitlength * \real{\svgscale}}%
    \fi%
  \else%
    \setlength{\unitlength}{\svgwidth}%
  \fi%
  \global\let\svgwidth\undefined%
  \global\let\svgscale\undefined%
  \makeatother%
  \begin{picture}(1,0.31345685)%
    \put(0,0){\includegraphics[width=\unitlength]{tempsnew.pdf}}%
    \put(0.31540208,0.3073761){\color[rgb]{0,0,0}\makebox(0,0)[rb]{\smash{$M/\mu=5/4$}}}%
    \put(0.83905994,0.30737621){\color[rgb]{0,0,0}\makebox(0,0)[rb]{\smash{$M/\mu=4.3/3$}}}%
    \put(0.56644273,0.05774685){\color[rgb]{0,0,0}\makebox(0,0)[rb]{\smash{$\sigma_{DC}$}}}%
    \put(0.99884479,0.05641063){\color[rgb]{0,0,0}\makebox(0,0)[rb]{\smash{$\rho$}}}%
    \put(0.05208992,0.29276568){\color[rgb]{0,0,0}\makebox(0,0)[rb]{\smash{$\sigma_{DC}$}}}%
    \put(0.49186494,0.2895166){\color[rgb]{0,0,0}\makebox(0,0)[rb]{\smash{$\rho$}}}%
    \put(0.04442039,0.25165234){\color[rgb]{0,0,0}\makebox(0,0)[rb]{\smash{1.6}}}%
    \put(0.04442039,0.18788052){\color[rgb]{0,0,0}\makebox(0,0)[rb]{\smash{1.4}}}%
    \put(0.04442039,0.12413688){\color[rgb]{0,0,0}\makebox(0,0)[rb]{\smash{1.2}}}%
    \put(0.04442039,0.06048754){\color[rgb]{0,0,0}\makebox(0,0)[rb]{\smash{1.0}}}%
    \put(0.11684182,0.02582522){\color[rgb]{0,0,0}\makebox(0,0)[rb]{\smash{0.5}}}%
    \put(0.1935503,0.02582522){\color[rgb]{0,0,0}\makebox(0,0)[rb]{\smash{1.0}}}%
    \put(0.27043482,0.02582522){\color[rgb]{0,0,0}\makebox(0,0)[rb]{\smash{1.5}}}%
    \put(0.34752681,0.02582522){\color[rgb]{0,0,0}\makebox(0,0)[rb]{\smash{2.0}}}%
    \put(0.42441135,0.02582522){\color[rgb]{0,0,0}\makebox(0,0)[rb]{\smash{2.5}}}%
    \put(0.27683123,0.00080139){\color[rgb]{0,0,0}\makebox(0,0)[rb]{\smash{$1/\mu$}}}%
    \put(0.49229501,0.25162415){\color[rgb]{0,0,0}\makebox(0,0)[rb]{\smash{2.7}}}%
    \put(0.49167595,0.1879087){\color[rgb]{0,0,0}\makebox(0,0)[rb]{\smash{1.8}}}%
    \put(0.49220005,0.12416507){\color[rgb]{0,0,0}\makebox(0,0)[rb]{\smash{0.9}}}%
    \put(0.49220005,0.06042141){\color[rgb]{0,0,0}\makebox(0,0)[rb]{\smash{0.0}}}%
    \put(0.79455093,0.00080156){\color[rgb]{0,0,0}\makebox(0,0)[rb]{\smash{$1/\mu$}}}%
    \put(0.61016703,0.02582511){\color[rgb]{0,0,0}\makebox(0,0)[rb]{\smash{0.5}}}%
    \put(0.68687552,0.02582511){\color[rgb]{0,0,0}\makebox(0,0)[rb]{\smash{1.0}}}%
    \put(0.76376004,0.02582511){\color[rgb]{0,0,0}\makebox(0,0)[rb]{\smash{1.5}}}%
    \put(0.84085203,0.02582511){\color[rgb]{0,0,0}\makebox(0,0)[rb]{\smash{2.0}}}%
    \put(0.91773657,0.02582511){\color[rgb]{0,0,0}\makebox(0,0)[rb]{\smash{2.5}}}%
    \put(0.55656418,0.2905911){\color[rgb]{0,0,0}\makebox(0,0)[rb]{\smash{1.0}}}%
    \put(0.55708819,0.22355138){\color[rgb]{0,0,0}\makebox(0,0)[rb]{\smash{0.9}}}%
    \put(0.55708819,0.15657972){\color[rgb]{0,0,0}\makebox(0,0)[rb]{\smash{0.8}}}%
    \put(0.55708819,0.08960873){\color[rgb]{0,0,0}\makebox(0,0)[rb]{\smash{0.7}}}%
    \put(0.99982151,0.29052304){\color[rgb]{0,0,0}\makebox(0,0)[rb]{\smash{1.9}}}%
    \put(0.99982151,0.22355517){\color[rgb]{0,0,0}\makebox(0,0)[rb]{\smash{1.3}}}%
    \put(1.00034564,0.15657972){\color[rgb]{0,0,0}\makebox(0,0)[rb]{\smash{0.8}}}%
    \put(1.00034564,0.08960873){\color[rgb]{0,0,0}\makebox(0,0)[rb]{\smash{0.2}}}%
  \end{picture}%
\endgroup%